\documentclass[a4paper,11pt]{article}
\pdfoutput=1 

\usepackage{jheppub} 

\usepackage[T1]{fontenc} 

\usepackage[usenames]{color}
\usepackage[dvipsnames]{xcolor}
\usepackage{placeins}

\newcommand{\cosmoconstant}{\ensuremath{\Lambda}} 
\newcommand{\newtoncoupling}{\ensuremath{G}} 
\newcommand{\betagauge}{\ensuremath{\beta_h}} 

\newcommand{\HyperC}{\ensuremath{g_{Y}}} 
\newcommand{\wA}{\ensuremath{w_2}} 
\newcommand{\vA}{\ensuremath{v_2}} 
\newcommand{\wAast}{\ensuremath{w_{2\, \ast}}} 
\newcommand{\vAast}{\ensuremath{v_{2\, \ast}}}

\newcommand{\kA}{\ensuremath{h_2}} 

\newcommand{\wB}{\ensuremath{x_2}} 
\newcommand{\vB}{\ensuremath{u_2}} 

\newcommand{\wAB}{\ensuremath{y_{2}}} 
\newcommand{\vAB}{\ensuremath{t_{2}}} 

\newcommand{\wABT}{\ensuremath{z_{2}}} 
\newcommand{\vABT}{\ensuremath{s_{2}}} 

\newcommand{\w}{\ensuremath{w_2}} 
\newcommand{\wT}{\ensuremath{y_2}} 

\newcommand{\vt}{\ensuremath{v_2}} 
\newcommand{\vT}{\ensuremath{t_2}} 

\newcommand{\ZA}{\ensuremath{Z_A}} 
\newcommand{\NV}{\ensuremath{N_{\mathrm{V}}}} 

\newcommand{\Int}{\int\!\!\mathrm{d}^4x\sqrt{g}\,} 
\newcommand{\Intd}{\int\!\!\mathrm{d}^dx\sqrt{g}\,} 
\newcommand{\Intb}{\int\!\!\mathrm{d}^4x\sqrt{\bar{g}}\,} 

\newcommand{\Gcrit}{{\newtoncoupling}_{\mathrm{crit}}} 

\newcommand{\OT}{\ensuremath{O(2)}}

\title{
The weak-gravity bound in asymptotically safe gravity-gauge systems
}

\author[a]{A. Eichhorn,}
\author[a,b]{J.H. Kwapisz,}
\author[a,c,d]{M. Schiffer}

\affiliation[a]{CP3-Origins, University of Southern Denmark, Campusvej 55, DK-5230 Odense M, Denmark}
\affiliation[b]{Institute of Theoretical Physics, Faculty of Physics, University of Warsaw, ul. Pasteura 5, Warsaw, Poland}
\affiliation[c]{Institut f\"ur Theoretische Physik, Universit\"at Heidelberg, Philosophenweg 16, 69120 Heidelberg, Germany}
\affiliation[d]{Perimeter Institute for Theoretical Physics, 31 Caroline St N, Waterloo ON, N2L 2Y5, Canada}

\emailAdd{eichhorn@cp3.sdu.dk}
\emailAdd{jkwapisz@fuw.edu.pl}
\emailAdd{mschiffer@perimeterinstitute.ca}

\abstract{
 The weak-gravity bound has been discovered in asymptotically safe gravity-matter systems, where it limits the maximum strength of gravitational fluctuations. In the present paper, we explore it for the first time in systems with more than one gauge field, to discover whether systems with 12 gauge fields (like the Standard Model) exhibit a weak-gravity bound and whether the gravitational fixed point evades it.\\
Further, we test the robustness of the present and previous results on the weak-gravity bound by exploring  their dependence on a gravitational gauge parameter.\\
Finally, the existence of the weak-gravity bound also has important phenomenological consequences: it is key to a proposed mechanism that bounds the spacetime dimensionality from above to four or five dimensions. In this paper, we strengthen the evidence for this mechanism. Thus, the predictive power of the asymptotic safety paradigm could extend to parameters of the spacetime geometry, such that the four-dimensionality of our universe could be explained from first principles.
}

\begin{document} 
\maketitle
\flushbottom

\section{Introduction}
\label{sec:intro}
How much can be learned about quantum gravity without direct observational access to Planck-scale phenomena? This question drives a significant part of research on the phenomenology of quantum gravity, see \cite{Addazi:2021xuf} for an extensive review. Within asymptotically safe quantum gravity (see \cite{Eichhorn:2017egq,Percacci:2017fkn,Eichhorn:2018yfc, Reuter:2019byg,Pereira:2019dbn,Reichert:2020mja,Pawlowski:2020qer} for reviews  and \cite{Bonanno:2020bil} for a critical discussion of the current status), it has been found that the existence of Standard Model (SM) matter at low energies strongly constrains the properties of the regime at transplanckian energies, in the following ways:\\
 First, the interaction structure of matter models at low energies constrains asymptotically safe gravity at high scales: to accommodate nonvanishing Yukawa couplings (giving rise to fermion masses in the SM), the values of gravitational couplings at transplanckian energy scales are constrained \cite{Eichhorn:2016esv, Eichhorn:2017eht, Eichhorn:2017ylw}. Similarly, specific beyond SM (BSM) models impose their own constraints on asymptotically safe gravity, see, e.g., \cite{DeBrito:2019rrh,Eichhorn:2020sbo}, while in turn the consistent coupling to asymptotically safe gravity reduces the parameter space in many BSM settings, see, e.g., \cite{Eichhorn:2017als, Grabowski:2018fjj, Eichhorn:2019dhg,Kwapisz:2019wrl, Reichert:2019car, Hamada:2020vnf, Kowalska:2020gie, Eichhorn:2020kca, Kowalska:2020zve}.\\
 Second, the existence of certain matter degrees of freedom is already enough to constrain asymptotically safe gravity, without considerations of specific interaction structures at low energies: if asymptotically safe gravity is too strongly coupled, the strong gravitational fluctuations destroy fixed points in matter interactions \cite{Eichhorn:2012va, Eichhorn:2016esv, Christiansen:2017gtg}. The matter interactions in question are not relevant for low-energy phenomenology, because they are canonically higher-order interactions which are induced by gravity \cite{Eichhorn:2012va,Eichhorn:2016esv,Christiansen:2017gtg,Eichhorn:2017eht,Eichhorn:2019yzm,deBrito:2021pyi,Laporte:2021kyp,deBrito:2020dta}. It is required that they lie at an interacting, i.e., asymptotically safe fixed point \footnote{For symmetry reasons, which we also review in the present paper, a non-interacting, i.e., asymptotically free fixed point is unavailable for these interactions.} at high energies and thus one can conclude that asymptotically safe gravity is restricted to be sufficiently weakly interacting. Thus, asymptotic safety has to satisfy the \emph{weak-gravity bound} (WGB)\footnote{Though similar in name, weak gravity bound should not be confused with weak gravity conjecture \cite{Arkani-Hamed:2006emk}. For the relation between string theory, asymptotic safety and the weak gravity conjecture, see \cite{deAlwis:2019aud, Basile:2021krr}.}.\\
This program of relating effective theories for matter to asymptotically safe gravity is in its aim partially similar to the swampland program in string theory  \cite{Vafa:2005ui}, see \cite{Brennan:2017rbf,Palti:2019pca} for reviews, in that it determines which effective field theories could be ultraviolet completed by their coupling to quantum gravity. However, it goes beyond the swampland program in that within asymptotic safety, these considerations narrow down the gravitational parameter space very significantly.\\
Both types of constraints on the transplanckian regime give rise to boundaries in the microscopic gravitational parameter space. These have been determined, but are subject to systematic uncertainties. Therefore, it is critical to extend previous work to determine those boundaries with reduced systematic uncertainties, in order to decide whether or not asymptotically safe gravity is phenomenologically viable. In this paper, we make a significant step in that direction with a focus on gauge interactions. First, we extend the work in \cite{Christiansen:2017gtg,Eichhorn:2019yzm} to a complete basis of interactions at a given order in fields and derivatives. Second, we consider settings with more than one gauge field and explore the dependence of the 
WGB on the number of vectors, similar to a corresponding recent study for scalar fields \cite{deBrito:2021pyi}. Third, we use gauge dependence as a proxy for the stability of results, i.e., investigate whether physical statements, such as the existence of a fixed point, are gauge independent, as they should \footnote{Within asymptotically safe gravity, approximations can result in a gauge dependence even in physical quantities. In turn, gauge dependence allows to determine whether approximations are robust (when gauge dependence is low/absent) or insufficient (when gauge dependence becomes large).}.\\
This paper is organized as follows. In Sec.~\ref{sec:weakgravity} we discuss the constraints a UV-complete Abelian gauge sector imposes on the gravitational parameter space in more detail. Sec.~\ref{sec:onegauge} is dedicated to the study of the system consisting of one field,  in Sec.~\ref{sec:twogauge}, two gauge fields are considered. In  sec.~\ref{sec:Ngauge}  we perform a study of $N_V >2$ gauge fields,  also investigating the large $N_V$ limit. Finally, in Sec.~\ref{sec:gaugedependence} we discuss the gauge dependence of the WGB as a test for the robustness of our results. Sec.~\ref{sec:discussion} summarizes our work. In App.~\ref{sec:WGBbetazero} we discuss the case when the WGB ceases to be a function in the space of couplings.
\section{The Abelian gauge sector in asymptotically safe quantum gravity}
\label{sec:weakgravity}
We will briefly review the status of the Abelian gauge sector in asymptotically safe quantum gravity. We will first focus on the Abelian gauge coupling $\HyperC$, and then discuss the role of higher-order, induced matter interactions.
\subsection{The Abelian hypercharge}
\label{sec:hyp}
Quantum fluctuations of charged matter have a screening effect on the Abelian gauge coupling, resulting in a Landau pole in perturbation theory and a non-perturbative triviality problem \cite{Gockeler:1997dn, Gockeler:1997kt, Gies:2004hy}. The associated scale of new physics is transplanckian, suggesting that the new physics required to solve the triviality problem could be quantum gravity.

For asymptotically safe gravity, the leading-order terms in the beta function for the Abelian hypercharge coupling, including the gravitational contribution, read
\begin{equation}
\label{eq:BetaGaugeGrav}
\beta_{\HyperC} = - f_g \, \HyperC + \frac{\HyperC^3}{16\pi^2} \frac{41}{6}+ \dots\,,
\end{equation}
where $f_g$ parameterizes the quantum gravitational contribution, which depends on the gravitational couplings, see, e.g., \cite{Eichhorn:2017lry} for the explicit form. 

Explicit computations using the functional Renormalization Group (FRG), cf.~\ref{sec:Setup},  yield  $f_g>0$ \cite{Daum:2009dn, Harst:2011zx,
Eichhorn:2017lry, Christiansen:2017gtg, Christiansen:2017cxa, Eichhorn:2018whv, Eichhorn:2019yzm} (with a general argument that $f_g\geq0$ in \cite{Folkerts:2011jz}), indicating an antiscreening effect of gravitational fluctuations \footnote{Since the gravitational couplings are not marginal, the gravitational contribution $f_g$ is not universal , but is scheme-dependent. Studies using dimensional regularization within perturbation theory indicate that $f_g=0$ \cite{Pietrykowski:2006xy, Toms:2007sk, Ebert:2007gf, Toms:2010vy, Anber:2010uj}, but neglect the contribution of higher-order couplings to the scale dependence of \HyperC \cite{Christiansen:2017gtg}, cf.~Sec.~\ref{sec:IndC}.
In addition, perturbative studies do not compute universal quantities; in contrast to the FRG setting, in which $f_g$, when evaluated on a gravitational fixed point, corresponds to  the critical exponent of the free fixed point ${\HyperC}_{*}=0$, and is thus universal. In the following, we will choose a scheme where $f_g>0$, and investigate scheme independent quantities, e.g., critical exponents or the existence of fixed points, in this scheme.}.
Thus, gravitational fluctuations change the scaling dimension of the gauge coupling near $\HyperC =0$, thereby solving the Landau pole/triviality problem. 

Additionally, gravitational fluctuations induce a second, interacting fixed point ${\HyperC}_{\, *}>0$, at which the gauge coupling is irrelevant, which means that its value at all scales is predictable. Thus, its infrared (IR) value can be computed from first principles, see \cite{Harst:2011zx,Eichhorn:2017lry}. Within approximations that cause significant systematic uncertainties, the calculated IR value is 35 $\%$ above the measured value. Accordingly, two universality classes are currently compatible with observations within the systematic uncertainties: First, from an asymptotically safe fixed point, the gauge coupling is predicted at all scales; second from an asymptotically free  fixed point the Abelian gauge coupling can reach any IR-value within an interval bounded by the prediction from the interacting fixed point. Given that the current estimate for this upper bound on the gauge coupling is 35 $\%$ above the measured value (with systematic uncertainties expected to be roughly of a similar size), asymptotically safe gravity appears to indeed solve the Abelian triviality problem.
\subsection{Higher order, induced interactions}
\label{sec:IndC}
Under the impact of gravity, additional matter interactions besides the couplings of the Standard Model have to be scrutinized, because these additional interactions are generated by gravitational fluctuations.
In particular, all matter interactions that are compatible with the symmetries of the kinetic term of a given matter field are expected to be induced by  gravitational fluctuations which are nonvanishing at an asymptotically safe fixed point \cite{Eichhorn:2017eht}. This general argument has been explicitly confirmed for scalars \cite{Eichhorn:2012va,Eichhorn:2013ug,Eichhorn:2017sok,deBrito:2021pyi,Laporte:2021kyp}, fermions \cite{Eichhorn:2011pc,Meibohm:2016mkp,Eichhorn:2018nda} and gauge fields \cite{Christiansen:2017gtg,Eichhorn:2019yzm}  as well as mixed scalar-fermion systems \cite{Eichhorn:2016esv,Eichhorn:2017eht}. For gauge fields, for instance the interaction $ \wA\,(F_{\mu\nu}F^{\mu\nu})^2$ is induced. Schematically, the scale dependence of $\wA$ at finite values of the dimensionless Newton coupling $\newtoncoupling$  reads
\begin{align}
\label{eq:betaw}
\beta_{\wA} = \wA\,C_1(\newtoncoupling) +C_0(\newtoncoupling)   + \wA^2\,C_2 +\dots\,,
\end{align}
where explicit expressions for $C_0, C_1$ and $C_2$ can be found in \cite{Christiansen:2017gtg}, and the ancillary notebook\footnote{The term linear in $\wA$ contains the canonical contribution generated by the dimensional nature of the $F_{\mu\nu}F^{\mu\nu}$ operator, and quantum contributions, i.e., $C_1=4+C_{1,\,\mathrm{q}}$}.
Their key property is that $C_0\to0$ as $\newtoncoupling\to0$. Consequently, in the absence of gravitational fluctuations, there is a free fixed point\footnote{ This applies to the setting of a pure gauge theory. In the presence of matter, matter loops also generate $\wA$, e.g., as part of the Euler-Heisenberg effective action. In contrast to the situation in gravity, the generation of $\wA$ by charged matter is an IR effect.} $\wAast=0$. Conversely, in the presence of gravitational fluctuations, $C_0\neq0$, such that $\wA=0$ is no longer a fixed point. Instead, the free fixed point is shifted into an interacting, shifted Gaussian fixed point (sGFP). At sufficiently weak gravitational interactions, the scaling exponent at the shifted Gaussian fixed point is close to the Gaussian one. The fixed-point value is given by
\begin{align}
\label{eq:FPSw2}
\wAast = \frac{-C_1(\newtoncoupling)}{2C_2} + \sqrt{\frac{C_1(\newtoncoupling)^2}{4C_2^2} - \frac{C_0(\newtoncoupling)}{C_2}}\,,
\end{align}
for $C_2>0$ \footnote{Actually, there are two fixed points differing by the sign in front of the square root. The sign of $C_2$ determines which one is the sGFP. In the following we assume that $C_2>0$, which we confirm by an
explicit computation later on.}. Hence, in the presence of gravity the gauge sector is necessarily interacting at high energies \cite{Eichhorn:2017eht,Christiansen:2017gtg,Eichhorn:2019yzm} and complete asymptotic freedom in the gauge sector cannot be achieved.  Instead, if a fixed point exists, it necessarily features a set of nonvanishing interactions.\\
Additionally, it is not even clear whether the gauge sector actually is UV complete under the impact of gravity, as suggested by studies focusing on the gauge coupling $\HyperC$ alone: 
The sGFP  becomes complex (and thus no longer a viable fixed point) due to a fixed-point collision at 
\begin{align}
\label{eq:WGBcondition}
C_{0,\,\mathrm{crit}}(\newtoncoupling)= \frac{C_1^2(\newtoncoupling)}{4C_2}\,.
\end{align} 
Therefore, for $C_0>C_{0,\,\mathrm{crit}}$, there is no real fixed point for $\wA$ and the gauge sector is not UV-complete for $C_0>C_{0,\,\mathrm{crit}}$. The condition on $C_0$ translates into a bound on the gravitational couplings. This bound is known as the \textit{weak gravity bound} (WGB) in the literature,  and appears in different gravity-matter systems \cite{Eichhorn:2016esv,Eichhorn:2017eht,Christiansen:2017gtg,Eichhorn:2019yzm,deBrito:2021pyi} at roughly similar values of the gravitational interactions \cite{Schiffer:2021gwl}. It owes its name to the fact that gravity has to be sufficiently weak in order for the sGFP to be real. Thus,
the WGB separates the region in the gravitational parameterspace, where a UV-completion of the matter sector is possible, from the excluded, strong-gravity region.

In summary, a UV-completion for the Abelian gauge sector requires that two conditions hold simultaneously, namely
\begin{align}
\label{eq:wgb}\centering
f_g>0 \quad \textrm{and} \quad C_0(\newtoncoupling)\leq C_{0,\,\mathrm{crit}}(\newtoncoupling).
\end{align}
The largest part of the gravitational parameter space satisfies both conditions in four dimensions, but no longer above five dimensions \cite{Eichhorn:2019yzm}, providing a constraint on the dimensionality of asymptotically safe gravity with Standard Model matter.

In the following, we extend the studies in \cite{Christiansen:2017gtg,Eichhorn:2019yzm} in various directions. In particular, we will investigate:
\begin{itemize}
\item  how a second gauge invariant interaction involving four gauge fields, given by $(F\tilde{F})^2$, impacts the WGB,
\item how this second gauge invariant interaction impacts the presence and value of a critical spacetime dimensionality, discovered in \cite{Eichhorn:2019yzm},
\item  which interaction structures are induced in a system containing $N_{\mathrm{V}}$ gauge fields,
\item how robust the WGB is under changes of the gauge choice, and number of gauge fields $N_{\mathrm{V}}$.
\end{itemize}

%
\section{One species of gauge fields}

\label{sec:onegauge}
\subsection{Setup}
\label{sec:Setup}
We use functional RG (FRG) techniques \cite{Wetterich:1992yh,Ellwanger:1993mw,Morris:1993qb} adapted to gravity \cite{Reuter:1996cp} to extract the scale dependence of the couplings and wavefunction renormalizations, see \cite{Berges:2000ew,Pawlowski:2005xe,Gies:2006wv,Delamotte:2007pf,Rosten:2010vm,Braun:2011pp,Reuter:2012id,Dupuis:2020fhh, Reichert:2020mja} for introductions and reviews. The FRG  realizes the Wilsonian  paradigm of integrating out  quantum fluctuations in the path integral according to their momentum shell. This is implemented for the scale-dependent effective action $\Gamma_k$, which interpolates between the classical action $S$ and the full effective action $\Gamma$.  The power and versatility of FRG techniques, see \cite{Dupuis:2020fhh} for a recent review from statistical physics to quantum gravity, comes from the functional differential equation for $\Gamma_k$. This flow equation encodes the change of $\Gamma_k$ in response to quantum fluctuations in the momentum shell between $k$ and $k-\delta k$. The flow equation reads
\begin{align}
\label{eq:Wetterich}
k\partial_k \Gamma_k = \frac{1}{2}\textrm{STr} \left[(\Gamma_k^{(2)} + R_k)^{(-1)}\,k\partial_k R_k \right],
\end{align}
where the matrix $\Gamma_k^{(2)}$ is the second functional derivative of $\Gamma_k$ with respect to the fields and also carries spacetime and internal indices. The super-trace STr involves the summation over these spacetime and internal indices, as well as an integration over $d$-dimensional space. 
The beta-functions of couplings, which parameterize the different interaction monomials in $\Gamma_k$ can be extracted from the flow equation by projecting onto the corresponding field monomials.The flow equation Eq.~\eqref{eq:Wetterich} can thus be understood as a compact summary of all beta functions and anomalous dimensions of the theory.

The regulator $R_k$ suppresses modes of the field with momenta below $k$ in the scale-dependent effective action $\Gamma_k$. The introduction of the regulator $R_k$ guarantees UV and IR finiteness of the flow equation, and implements the momentum-shell wise integration of quantum fluctuations.
To ensure this, the regulator function has to satisfy several conditions, but a certain freedom in its choice exists. We use this freedom to select a regulator which is proportional to the momentum-dependent part of $\Gamma_k^{(2)}$, i.e.,
\begin{equation}
R_k=\Gamma_k^{(2)}\Big|_{\cosmoconstant=0}\,\, r_k\left(p^2/k^2\right)\,,
\end{equation}
with the shape-function $r_k$. This spectrally adjusted regulator ensures that no further symmetries are broken by the regulator, see \cite{Gies:2002af, Pawlowski:2005xe, Benedetti:2010nr, Gies:2015tca}.For the shape-function $r_k$, we choose a Litim-type cutoff \cite{Litim:2001up}
\begin{align}
r_k(p^2/k^2) = \left(\frac{k^2}{p^2}-1 \right)\theta\left(1-\frac{p^2}{k^2}\right),
\end{align}
which gives rise to analytic expressions for the beta-functions. 

Even though along the flow all field monomials compatible with the symmetries are induced, for practical calculations one has to truncate the effective action. We approximate the dynamics of the system by 
\begin{equation}
\Gamma_k=\Gamma_k^{\mathrm{EH}}+\Gamma_k^{U(1)}\,,
\end{equation}
with the Einstein-Hilbert action
\begin{equation}
\label{eq:Gammaktot}
\Gamma_k^{\mathrm{EH}}=-\frac{1}{16 \pi \newtoncoupling k^{-2}} \Int (R-2  \cosmoconstant k^2) + S_{\mathrm{gf},\, h}\,,
\end{equation}
where we have introduced the dimensionless versions of the Newton coupling $\newtoncoupling$  and of the cosmological constant $\cosmoconstant$, respectively. To include gravitational fluctuations, we apply the background-field method, and decompose the full metric into a background metric and a fluctuation field, according to
\begin{align}
\label{eq:linearsplit}
g_{\mu\nu} = \delta_{\mu\nu} + h_{\mu\nu}\,.
\end{align}
We  choose a flat background which suffices to extract all beta functions that we are interested in. To calculate the propagator for the fluctuation field, we introduce a gauge-fixing condition $\mathcal{F}^{\mu}=0$, with 
\begin{equation}
\mathcal{F}^{\mu} = \left(\delta^{\mu\kappa}\bar{D}^{\lambda} - \frac{1+ \betagauge}{4} \delta^{\kappa\lambda}\bar{D}^{\mu}\right) h_{\kappa\lambda}\,.
\end{equation}
$\betagauge$ is one of two gauge parameters, which we fix to $\beta_h=1$ in the following, except for Sec.~\ref{sec:gaugedependence}, where we keep it more general. The second, $\alpha_h$, enters the gauge-fixing action
\begin{align}
S_{\mathrm{gf},\, h} &= \frac{1}{32\,\pi\, \alpha_h\, \newtoncoupling k^{-2}} \Intb\mathcal{F}^{\mu} \bar{g}_{\mu\nu}\mathcal{F}^{\nu}\,, \quad \alpha_h \to 0\,.
\end{align}
The gauge fixing also introduces Fadeev-Popov ghosts. These might contribute to the beta functions for the Abelian gauge field indirectly, through  induced ghost-matter interactions \cite{Eichhorn:2013ug} which we neglect here.

For the Abelian gauge sector, we approximate the dynamics by the standard kinetic term, and the two independent and gauge invariant interactions  at mass dimension eight for four gauge fields:
\begin{align}
\label{eq:gammakone}
\Gamma_k^{U(1)} =& \frac{1}{4} \Int  F_{\mu\nu} F^{\mu \nu}\,+\,S_{\mathrm{gf},\, A}   \nonumber\\
& + \frac{\wA \,k^{-4}}{8}  \Int ( F_{\mu\nu} F^{\mu\nu})^2  
+ \frac{\vA \,k^{-4}}{8}  \Int
( F_{\mu\nu} \tilde{F}^{\mu\nu})^2\,.
\end{align}
$F_{\mu\nu}= \partial_{\mu} A_{\nu} - \partial_{\nu} A_{\mu}$ is the field strength tensor, and $\tilde{F}_{\mu\nu} = \frac{1}{2} \epsilon_{\mu\nu\rho\sigma}F^{\rho\sigma}$ its dual tensor, with the totally antisymmetric tensor $\epsilon_{\mu\nu\rho\sigma}$. In the presence of a non-flat metric, $\epsilon_{\mu\nu\rho\sigma}$ is related to its flat space counterpart $\epsilon_{ijkl}$ by $\sqrt{g}^{-1}$. The interactions labeled by the dimensionless and scale-dependent couplings $\wA$ and $\vA$ are the  only two independent and gauge invariant interactions up to this order in mass dimension \footnote{ We neglect the operators $F_{\mu\nu}\square F^{\mu\nu}$ and $F_{\mu\nu}\square^2 F^{\mu\nu}$ here, which would contribute to the momentum-dependent anomalous dimension of the gauge field, see \cite{Knorr:2021slg}.}. The action $\Gamma_k^{U(1)}$ defined in Eq.~\eqref{eq:gammakone} actually contains the same terms as the weak-field expansion of the Euler-Heisenberg effective action \cite{Heisenberg:1936nmg}. The gauge-fixing action for the $U(1)$ gauge field is implemented by
\begin{equation}
S_{\mathrm{gf},\,A} = \frac{1}{2\alpha_A} \Intb\left(\bar{D}^{\mu}A_{\mu}\right)\left(\bar{D}^{\mu} A_{\mu}\right)\,,\quad \alpha_A\to 0\,,
\end{equation}
and the corresponding Fadeev-Popov ghosts decouple from all beta functions in the non-gravitational sector.

After expanding the action Eq.~\eqref{eq:Gammaktot} to second order in metric fluctuations $h_{\mu\nu}$, we perform a rescaling of the fields to bring the kinetic terms to canonical form:
\begin{align}
\label{eq:resc}
h_{\mu\nu} \to \sqrt{Z_h 16\pi k^{-2}\newtoncoupling}\, h_{\mu\nu}\,,\quad \text{and}\quad A_{\mu}\to \sqrt{Z_A}\,A_{\mu}\,.
\end{align}
The wave-function renormalizations $Z_h$ and $Z_A$ for metric fluctuations and the gauge field absorb the scale-dependence of the respective field \footnote{ In the presence of diffeomorphism symmetry breaking by the regulator and gauge fixing, there is more than one ``avatar" of the Newton coupling. However,
with  our rescaling of the metric fluctuation we introduce one single gravitational coupling $\newtoncoupling$ for each of the distinct gravity-matter vertices, as well as a single graviton-mass parameter. These choices assume a near-perturbative nature of quantum gravity, where the modified Slavnov-Taylor identities, which relate different gravity-matter and pure-gravity vertices, are trivial. Evidence for the agreement of different gravity-matter vertices has been found in \cite{Denz:2016qks, Eichhorn:2018akn, Eichhorn:2018nda, Eichhorn:2018ydy}, see also \cite{Knorr:2021niv} for a comparison of different graviton mass parameters. }. The wave-function renormalizations give rise to anomalous dimensions, defined as 
\begin{align}
\label{eq:anomalous}
\eta_h = - \partial_t \ln Z_h, \quad \eta_A = - \partial_t \ln Z_A\,.
\end{align} 

%
\subsection{Results}
\begin{figure}[tbp]
\includegraphics[width=.475\textwidth,clip]{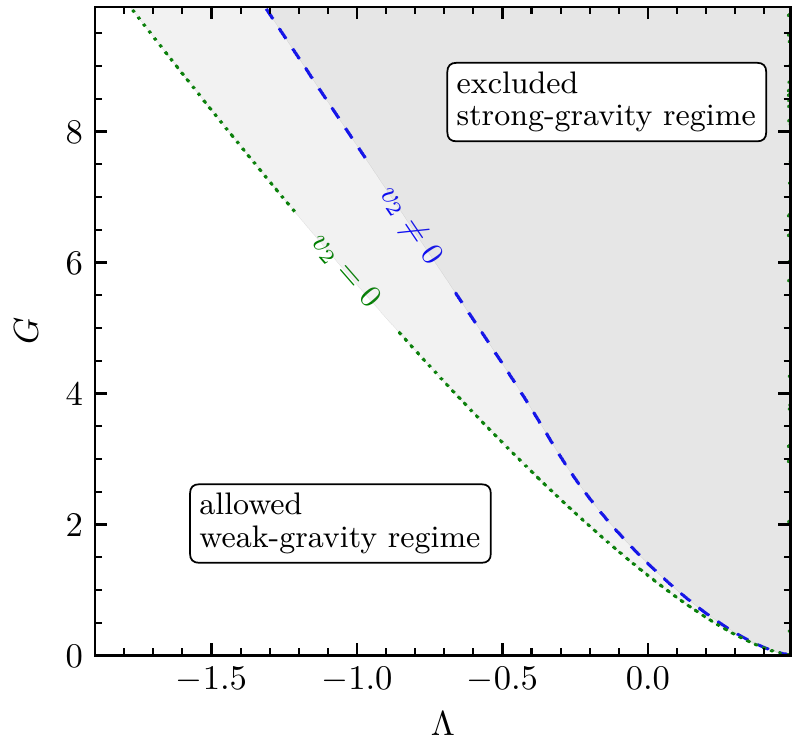}
\hfill
\includegraphics[width=.475\textwidth,clip]{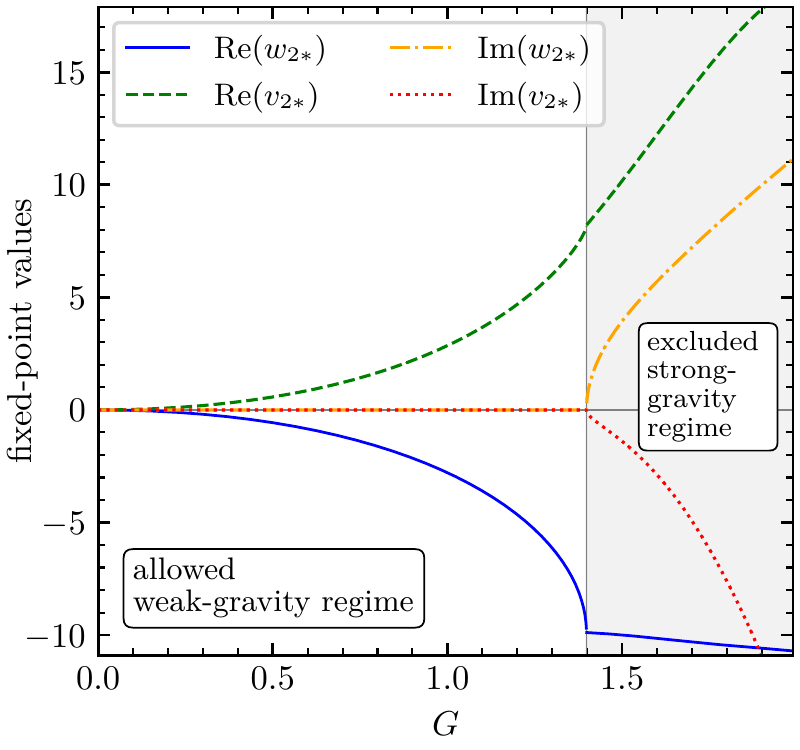}
\caption{
\label{fig:onegauge} 
Left panel: We show the WGB in the $\newtoncoupling-\Lambda$ plane, which separates the allowed weak-gravity regime (white) from the excluded strong-gravity regime (grey). The green dotted line shows the WGB from previous work, where $\vA$ was neglected; the blue dashed line shows the WGB under the inclusion of $\vA$ that is consistently evaluated at its fixed-point value $\vAast =\vAast(\newtoncoupling, \Lambda).$ \newline
Right panel: We show the real and imaginary parts of the two fixed-point values $\wAast$ and $\vAast$ as a function of $\newtoncoupling$ for $\Lambda=0$. The WGB is visible as the point at which the imaginary parts depart from zero.
}
\end{figure}

%
We now evaluate the scale dependence of the matter interactions $\wA$ and $\vA$, and of $\eta_A$. Their diagrammatic representation can be found in \cite{Christiansen:2017gtg, Eichhorn:2019yzm}, and we employ the Mathematica packages \textit{xAct} \cite{Brizuela:2008ra,Martin-Garcia:2007bqa,Martin-Garcia:2008yei, 2008CoPhC.179..597M, 2014CoPhC.185.1719N}, as well as the \textit{FormTracer} \cite{Cyrol:2016zqb}, for their evaluation.
For the purpose of simplicity, we will discuss the choice $\betagauge=1$ in the following.  We provide the analytical expressions at $\Lambda=0$ below, because these suffice to understand the key mechanisms at play. The full results can be found  in the  ancillary notebook.
In the perturbative approximation, where the anomalous dimension coming from the regulator insertion is neglected, the scale-dependences are given by
\begin{alignat}{3}
f_g&= -\frac{ \eta_A}{2}=
\frac{\newtoncoupling}{4 \pi} - \frac{\vA}{24\pi^2} - \frac{\wA}{6\pi^2},\hspace{-12pt}&&\hspace{24pt}& &\label{eq:fgfull}\\
\beta_{\wA} 
&=\left(4-\frac{7\,\newtoncoupling}{2\pi} + \frac{5\,\vA}{12\pi^2}\right)\wA\, && +\left( 8\newtoncoupling^2 - \frac{\newtoncoupling\,\vA}{\pi} + \frac{\vA^2}{6\pi^2}\right)  & &+ \frac{35}{24\pi^2}\wA^2\, , \label{eq:betaw2}\\
\beta_{\vA}
&=\left(4-\frac{25\,\newtoncoupling}{6\pi} + \frac{11\,\wA}{12\pi^2}\right)\vA\, &&-\left( 8\newtoncoupling^2+\frac{7\,\newtoncoupling\,\wA}{6\pi}+\frac{\wA^2}{24\pi^2}\right)   & & +\frac{1}{8\pi^2}\vA^2\,.\label{eq:betav2}
\end{alignat}
 We observe that, depending on the signs of $\vA$ and $\wA$, these contributions could counteract the gravitational antiscreening in Eq.~\eqref{eq:fgfull}. As it turns out $\vAast \approx - \wAast$ and $\wAast<0$, cf.~Fig.~\ref{fig:onegauge}, such that $f_g$ is antiscreening also in the presence of four-photon interactions.

We now explore whether or not the two Eqns.~\eqref{eq:betaw2} and \eqref{eq:betav2} give rise to a WGB. Because the system of equations is coupled, this is less straightforward than for a single beta function. We start with the observation that Eq.~\eqref{eq:betaw2} gives rise to a WGB for  small enough $\vA$. Conversely,  Eq.~\eqref{eq:betav2} does not give rise to a WGB for small enough $\wA$, because the signs of the quadratic term $\sim \vA^2$ and the gravitational term $\sim \newtoncoupling^2$ are opposite, cf.~\eqref{eq:FPSw2}. However, at large enough $\wA$, the term $\sim \wA^2$ in Eq.~\eqref{eq:betav2} might change the situation. Similarly, at large enough $\vA$, the term $\sim \newtoncoupling\, \vA$ in Eq.~\eqref{eq:betaw2} might prevent the WGB. In turn, the fixed-point values $\wAast$ and $\vAast$ grow in magnitude with $\newtoncoupling$. Unless they grow faster than the contributions $\sim \newtoncoupling$ in the beta functions, the conclusion holds that a WGB arises from Eq.~\eqref{eq:betaw2}, but no second bound from Eq.~\eqref{eq:betav2} independently. \\
A numerical study of fixed-point solutions confirms this conclusion: The left panel in Fig.~\ref{fig:onegauge} shows that the WGB is only shifted somewhat by the inclusion of $\vA$, but not altered qualitatively. Indeed, the fixed-point value for $\vA$ turns out to be positive, thus leading to a shift of the weak-gravity bound to larger $\newtoncoupling$ (at constant $\Lambda$).

The signs of the fixed-point values $\wAast$ and $\vAast$ might tempt one to speculate about the stability-properties of the fixed point. We caution that for stability studies, the IR values of couplings are decisive, not the UV fixed-point values. Nevertheless, a negative sign for $\wA$ might indicate that in the UV, the effective potential $W(F^2) = F^2 + w_2 (F^2)^2+...$ is not locally stable about the origin, $F^2=0$, but no reliable conclusion can be drawn without a study of higher-order terms. We caution further that the sign of the fixed-point values for $\wA$ and $\vA$ can be changed by a change of basis, see Sec.~\ref{sec:DimCrit}, emphasizing that the full effective potential $V(F^2,\,F\tilde{F})$ is required to infer the presence of non-trivial minima.

\subsection{Critical dimensionality revisited}
\label{sec:DimCrit}
In the following we extend the study of \cite{Eichhorn:2019yzm} and investigate, how the second independent four-gauge-field interaction impacts the critical dimensionality  for asymptotic safety. The critical dimensionality arises from a mechanism  discovered in \cite{Eichhorn:2019yzm}:
In dimensions $d>4$, the Abelian  gauge coupling is dimensionful, i.e., $\bar{g}_Y=(4-d)/2$. This results in an additional contribution to the scale-dependence of the dimensionless  counterpart of the gauge coupling $\HyperC$, i.e.:
\begin{equation}
\beta_{\HyperC}=\left(\frac{d-4}{2}-f_g\right)\HyperC+\mathcal{O}(\HyperC^3)\,.
\end{equation}
In $d>4$, this dimensional contribution competes with the anti-screening contribution from metric fluctuations, because both take the form of a scaling dimension. 
The overall scaling dimension has to be positive for a UV completion, see Sec.~\ref{sec:weakgravity},
\begin{equation}
\label{eq:reqGCd}
f_g(d)>\frac{d-4}{2}\,,
\end{equation}
such that the Abelian gauge coupling is a relevant direction at the Gaussian fixed point. From Eq.~\eqref{eq:reqGCd}, it follows that the gravitational contribution $f_g$ would have to increase with the dimensionality $d$ to induce a UV-completion for all values of $d$. However, $f_g$ actually shows the opposite behavior and decreases with the dimensionality (for fixed values of the gravitational couplings), see \cite{Eichhorn:2019yzm}. Thus, the required growth of $f_g(d)$ with $d$ requires increasing values of the gravitational couplings, i.e., a shift into the strong-gravity regime.
However, the WGB for the induced four-gauge interactions excludes this regime from the viable parameter space. In summary, there are two competing effects: A sufficiently large $f_g$ requires an increasing strength of gravity with increasing $d$; avoiding the WGB imposes a bound on the strength of gravity that is present at all $d$.
The competition of these two effects gives rise to a critical dimensionality, beyond which the Abelian gauge sector remains UV-incomplete. In a truncation only taking into account the $\wA (F^2)^2$ interaction, this critical dimensionality was found to be $d_{\mathrm{c}}\approx5.8$ \cite{Eichhorn:2019yzm}.\\

\begin{figure}[tbp]\centering
	\includegraphics[width=.475\textwidth]{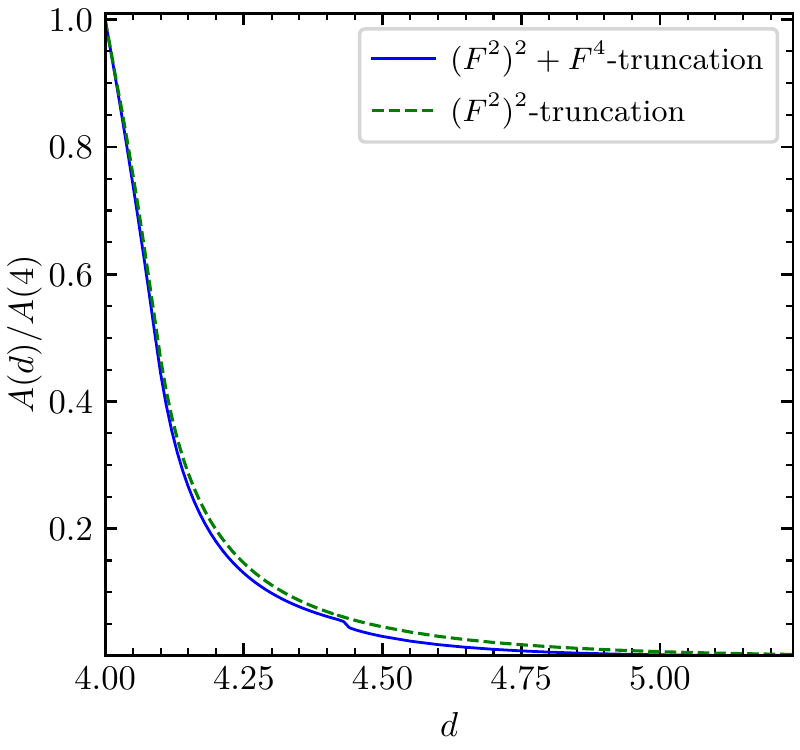}
	\caption{
		\label{fig:ADComp} 
		We show the area of the viable region in the gravitational paramaterspace for $\Lambda\in[-1500,0.5)$ and $\newtoncoupling\in[0,1000]$, where the Abelian gauge sector  could be UV-complete in the presence of gravitational fluctuations, as a function of the dimensionality $d$. The green (dashed) line shows the area of the viable region $A(d)$ in the pure $(F^2)^2$ truncation studied in \cite{Eichhorn:2019yzm}. The blue (solid) line shows the  area of the viable region $A(d)$  in the more complete truncation defined in Eq.~\eqref{eq:gammakoneN}. In both truncations, the viable region in the investigated parameterregion shrinks to zero at $d_{\mathrm{c}}<6$.
	}
\end{figure}

We test  how robust this result is by adding the second independent four-gauge-field interaction\footnote{A study of the gauge-dependence of the existence and value of the critical dimensionality as an independent test of the robustness of the system can be found in \cite{Schiffer:2021gwl}, showing only a quantitative impact of the gauge-parameter $\betagauge$ on $d_{\mathrm{c}}$ in the $\wA (F^2)^2$ truncation.}. Because the dual field-strength tensor $\tilde{F}_{\mu\nu}$ in Eq.~\eqref{eq:gammakone} is not a two-tensor away from $d=4$ dimensions, we exploit that in Euclidean spaces\footnote{We thank Benjamin Knorr for pointing this out to us.}
\begin{align}
F^{\mu\rho} F^{\nu}_{\phantom{\nu}\rho} + \tilde{F}^{\mu\rho} \tilde{F}^{\nu}_{\phantom{\nu}\rho}  &=\frac{1}{2}F^{\rho\sigma}F_{\rho\sigma}\, g^{\mu\nu}\,,\\
F^{\mu\rho}\tilde{F}^{\nu}_{\phantom{\nu}\rho} =\tilde{F}^{\mu\rho}F^{\nu}_{\phantom{\nu}\rho}& =\frac{1}{4}F^{\rho\sigma}\tilde{F}_{\rho\sigma}\,g^{\mu\nu}\,,
\end{align}
which holds in $d=4$, see \cite{Dittrich:2000zu, Knorr:2017kye} for the Lorentzian version, and from which it follows that
\begin{equation}
(F^{\rho\sigma}\tilde{F}_{\rho\sigma})(F^{\mu\nu}\tilde{F}_{\mu\nu}) =-4\,F^{\mu}_{\phantom{\mu}\nu}F^{\nu}_{\phantom{\nu}\rho}F^{\rho}_{\phantom{\rho}\sigma} F^{\sigma}_{\phantom{\sigma}\mu}+2(F^{\rho\sigma}F_{\rho\sigma})(F^{\mu\nu}F_{\mu\nu})\,,
\end{equation}
such that we can replace the $\vA (F\tilde{F})^2$ interaction in Eq.~\eqref{eq:gammakone} with a $\kA F^4$ interaction, because the difference between them is proportional to $w_2 (F^2)^2$. In $d=4$ dimensions, this is just a different choice of basis operators. The physical properties of the system, for example critical exponents, and the existence and location of a WGB  are the same in both bases,  as we have explicitly confirmed in our computation.  In $d>4$ dimensions, the $F^4$ basis is the appropriate one to work with.
To extend the study of a single species of gauge field to $d>4$, we generalize the matter part of the scale-dependent effective action in Eq.~\eqref{eq:gammakone} to
\begin{align}
\label{eq:gammakoneN}
\Gamma_k^{U(1)} =& \frac{1}{4} \Intd  F_{\mu\nu} F^{\mu \nu}\,+\,S_{\mathrm{gf},\, A}   \nonumber\\
& + \frac{\wA \,k^{-d}}{8}  \Intd ( F_{\mu\nu} F^{\mu\nu})^2  
+ \frac{\kA \,k^{-d}}{8}  \Intd
( F^{\mu}_{\phantom{\mu}\nu}F^{\nu}_{\phantom{\nu}\rho}F^{\rho}_{\phantom{\rho}\sigma} F^{\sigma}_{\phantom{\sigma}\mu})\,.
\end{align}
With this action, we can investigate the WGB for the couplings $\wA$ and $\kA$, as well as the condition \eqref{eq:reqGCd} for arbitrary dimensionality.  For a given dimensionality, these two conditions give rise to a region in parameter space, where metric fluctuations are strong enough to render the Abelian gauge coupling UV-complete, but weak enough to avoid the WGB for induced interactions. The two-dimensional area of that region (spanned by $\newtoncoupling \in [0,1000]$ and $\Lambda \in [-1500,0.5]$) is denoted $A(d)$ see \cite{Eichhorn:2019yzm} for details. 
In Fig.~\ref{fig:ADComp} we compare $A(d)$  in the $(F^2)^2$ truncation studied in \cite{Eichhorn:2019yzm} (green, dashed line) with the truncation defined in Eq.~\eqref{eq:gammakoneN} (blue, solid line). In both cases, we normalize to the allowed region for $d=4$. The inclusion of the second independent interaction $\kA$ clearly only has a  subleading quantitative effect. The main result of \cite{Eichhorn:2019yzm}, namely the existence of a critical dimension $d_{\mathrm{c}}<6$, above which the Abelian gauge sector remains UV-incomplete, even in the presence of quantum gravity, remains unchanged. In fact, the inclusion of $\kA$ reduces the critical dimensionality to $d_{\mathrm{c}}\approx 5.5$, compared to   $d_{\mathrm{c}}\approx 5.8$ reported in \cite{Eichhorn:2019yzm}.

We interpret this small impact of the second four-gauge-field interaction as an indication for the robustness of our study. Thus, four (and, with a significantly reduced parameter space also five) dimensions indeed appear to be special in asymptotically safe gravity-matter models. 

\section{Two species of gauge fields}
\label{sec:twogauge}
We now extend the system and add a second Abelian gauge field. 
The motivation for such an extension is twofold: First, nature might contain more than a single Abelian gauge field, e.g., dark photons are considered in Standard-Model extensions \cite{Fabbrichesi:2020wbt}. Second, gravitational effects are independent of internal symmetries, because gravity is ``blind" to those. Accordingly, our study of several Abelian gauge fields is also relevant for non-Abelian gauge groups, at least at small values of the non-Abelian gauge coupling, which remains asymptotically free in the presence of gravitational fluctuations \cite{Daum:2009dn,Folkerts:2011jz,Christiansen:2017cxa}, see also the corresponding discussion in \cite{Eichhorn:2017eht}.

We study a gauge field $A^{a}_{\mu}$, where $a\in(1,2)$  is a species-index, with field strength tensor $F^{a}_{\mu\nu}$. For a study of the gravity-induced interactions of these two gauge fields, a thorough understanding of their \emph{global} symmetries is critical. These global symmetries are always preserved by the RG flow, unless explicit symmetry violations are introduced by the regulator. Starting from the two kinetic terms, i.e., $F_{\mu\nu}^aF^{\mu\nu,\, a}$, we identify the following global symmetry: The sum of the two kinetic terms is preserved under the global $\OT$ rotation
\begin{align}
\label{eq:O2symmetry}
\left(\begin{array}{c} A^1_{\mu}\\ 
A^2_{\mu} \end{array}\right) \to \left( \begin{array}{cc} \cos \theta & \sin \theta \\
 -\sin \theta & \cos \theta \end{array}\right) \left(\begin{array}{c} A^1_{\mu}\\ A^2_{\mu}\end{array}\right).
\end{align} 
For the special case of $\theta=\pi$, this reduces to two independent $\mathbb{Z}_2$ symmetries, under which $A_{\mu}^{1} \rightarrow - A_{\mu}^1$, or $A_{\mu}^{2} \rightarrow - A_{\mu}^2$ respectively. These two $\mathbb{Z}_2$ symmetries are preserved even if the two gauge fields acquire different anomalous dimensions, i.e., if the rescaling \eqref{eq:resc} is performed for both gauge fields individually. If the anomalous dimensions are the same, then the full global $\OT$ symmetry is preserved by the flow.\\
It was recently investigated for scalar fields, whether the analogous global $\OT$ symmetry is preserved by the flow and a positive answer was found \cite{deBrito:2021pyi}. Together with an argument about the diagrammatic structure of the flow equation (spelled out in \cite{Eichhorn:2020mte}), where no symmetry-breaking terms are generated if regulator, propagator and interaction vertices satisfy a symmetry, we do not expect that violations of the $\OT$ symmetry are induced by gravitational fluctuations.  Nevertheless, we test this hypothesis explicitly, by working in an enlarged space of couplings, which only preserves the two $\mathbb{Z}_2$ symmetries. 

\begin{figure}[tbp]
\includegraphics[width=.475\textwidth,clip]{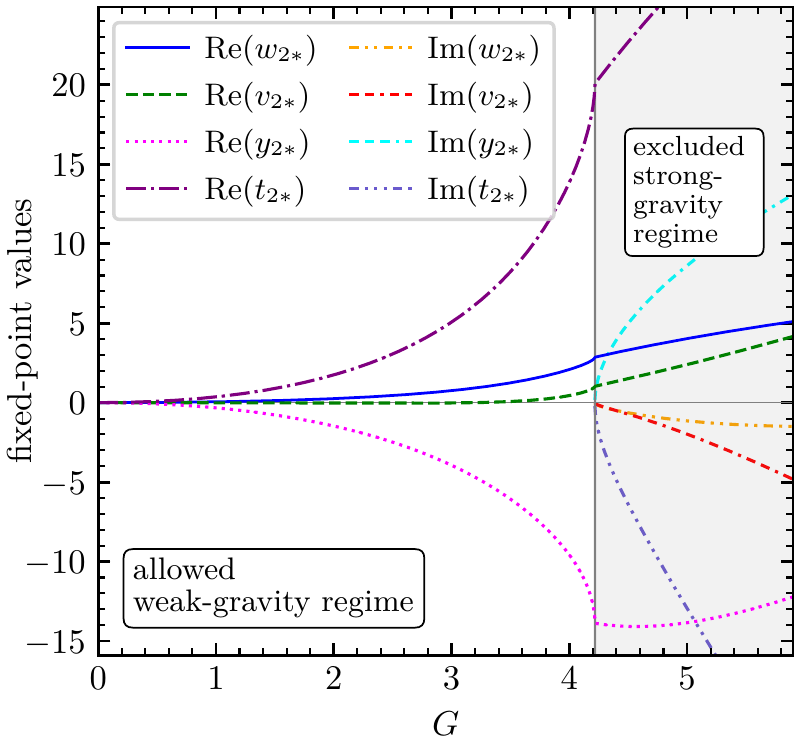}
\hfill
\includegraphics[width=.475\textwidth,clip]{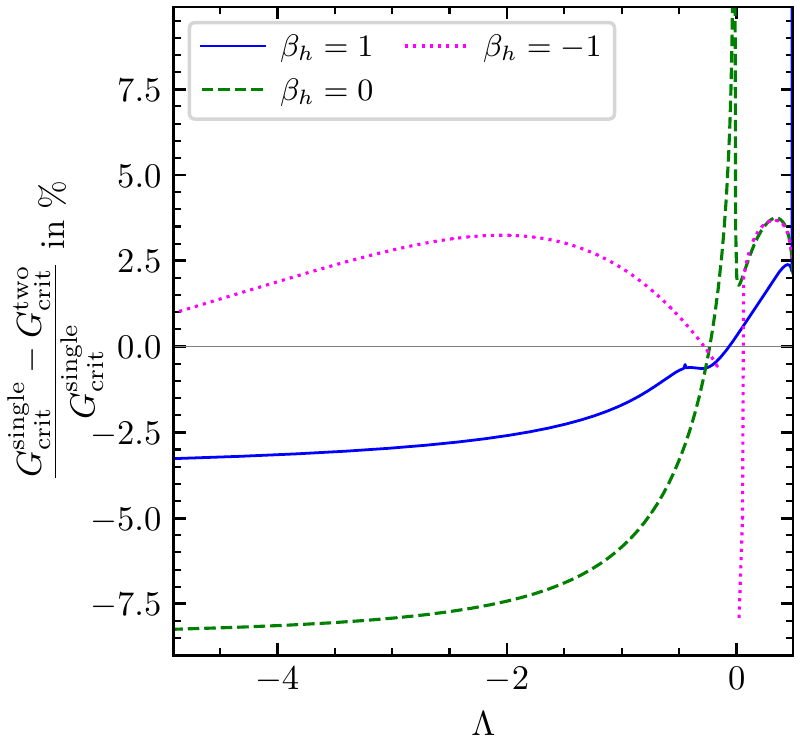}
\caption{
\label{fig:twogauge} 
Left panel: We show the real and imaginary parts of the four fixed-point values of the $\OT$-symmetric, two-species system described by Eq.~\eqref{eq:gammaktwogauge} as a function of $\newtoncoupling$ for $\Lambda=-0.5$. We show the imaginary parts of the sGFP, once the fixed-point values become complex. The WGB is visible as the point at which the imaginary parts depart from zero, marked by the vertical gray line.
Left panel: We show the relative deviation of the critical value of the Newton coupling $\Gcrit$ between the single-gauge field system (see Eq.~\eqref{eq:gammakone}), and the $\OT$ symmetric system of two vector fields (see Eq.~\eqref{eq:gammaktwogauge}). $\Gcrit$ is defined as the value for the Newton coupling, where the sGFP becomes complex, i.e., the value of the WGB as a function of $\Lambda$, see \eqref{eq:Gcritdef}. The blue (solid) line shows the comparison of both systems for $\betagauge=1$, indicating that the inclusion of the second gauge field only modifies the location of the WGB on the percent level. The green (dashed) and magenta (dotted) line show the same quantity for different choices of the gauge-fixing parameter $\betagauge$, see Sec.~\ref{sec:gaugedependence} for a discussion. We do not plot the magenta (dotted) line for $\Lambda\in(-0.16,0.02)$ due to numerical artefacts, which we briefly discuss in App.~\ref{sec:WGBbetazero}.
}
\end{figure}

The most general action for the gauge fields, which satisfies these symmetries, and including all linearly independent interaction up to dimension eight with four gauge fields is given by
\begin{align}
\label{eq:gammaktwogauge}
\Gamma_k^{U(1)\times U(1)} =& \frac{1}{4} \Int F^{a}_{\mu\nu} F^{\mu \nu,\,a} \,+\,S_{\mathrm{gf},\, A} \nonumber \\
&+ \frac{k^{-4}}{16}  \Int%
\bigg(\wA\,\left[F^a_{\mu\nu} F^{\mu\nu,\,a}  \right]^2+ \wAB\, (F^a_{\mu\nu} F^{\mu\nu,\,b})(F^a_{\rho\sigma} F^{\rho\sigma,\,b})\nonumber\\
&%
\hspace{75pt}+ \wB \,\left[ F^2_{\mu\nu} F^{\mu\nu,\,2}\right]^2+\wABT\,(F^1_{\mu\nu} F^{\mu\nu,\,2})(F^1_{\rho\sigma} F^{\rho\sigma,\,2})\bigg)\nonumber\\
&+ \frac{k^{-4}}{16}  \Int%
\bigg(\vA\,\left[ F^a_{\mu\nu} \tilde{F}^{\mu\nu,\,a}\right]^2+ \vAB\,(F^a_{\mu\nu} \tilde{F}^{\mu\nu,\,b})(F^a_{\rho\sigma} \tilde{F}^{\rho\sigma,\,b})\nonumber\\
&%
\hspace{75pt}+ \vB \,\left[F^2_{\mu\nu} \tilde{F}^{\mu\nu,\,2}\right]^2+\vABT\,( F^1_{\mu\nu} \tilde{F}^{\mu\nu,\,2})(F^1_{\rho\sigma} \tilde{F}^{\rho\sigma,\,2})\bigg),
\end{align}
where we have introduced the scale-dependent and dimensionless couplings $\wB$, $\wAB$, $\wABT$, $\vB$, $\vAB$, and $\vABT$, describing the self-interaction of the second gauge field, and the interaction between both fields. 
For these interactions, the $\OT$ symmetry of the kinetic term is only respected, if
\begin{align}
 \wB =0 , \quad \vB = 0, \quad \wABT = 0, \quad  \vABT = 0\,,
\end{align}
and for equal wave-function renormalizations of the two gauge fields.
Our explicit computations confirm that both anomalous dimensions are equal and
\begin{align}
\big(\beta_{\wB},\beta_{\wABT},\beta_{\vB},\beta_{\vABT}\big)\big|_{\wB=\wABT=\vB=\vABT=0}=(0,0,0,0)\,,
\end{align}
indicating that the $\OT$-symmetry breaking couplings $\wB$, $\wABT$, $\vB$, and $\vABT$ can be consistently set to zero.
Hence, once restricted to the $\OT$ theory-space, the non-$\OT$-symmetric interactions $\wB$, $\wABT$, $\vB$ and $\vABT$ are not induced by gravitational fluctuations. Based on this finding, we will restrict the following analysis to the $\OT$-symmetric theory-space. This reduces the number of induced couplings to $\wA$, $\vA$, $\wAB$, and $\vAB$, and a single wave-function renormalization $\ZA$. The beta functions for $\Lambda=0$ (for the general expression see the ancillary notebook) are given by
\begin{align}
\label{eq:twogaugebetas}
\beta_{\wA}=&\left(4-\frac{13 \newtoncoupling}{6 \pi }+\frac{5 \vAB}{16 \pi ^2}+\frac{5 \vA}{24 \pi ^2}+\frac{23 \wAB}{24 \pi ^2}\right)\,\wA +  \frac{55}{48 \pi ^2} \wA^2\nonumber\\
&+\newtoncoupling \left(\frac{ \vA}{2 \pi }-\frac{3\vAB}{4 \pi }-\frac{2  \wAB}{3 \pi }\right)+\frac{5 \vAB^2}{192 \pi ^2}+\frac{\vAB \vA}{48 \pi ^2}+\frac{\vAB \wAB}{12 \pi ^2}+\frac{\vA^2}{16 \pi ^2}+\frac{17 \wAB^2}{192 \pi ^2}\,,
\nonumber\\[4pt]
\beta_{\wAB} =& \left(4-\frac{17 \newtoncoupling}{6 \pi }+\frac{\vAB}{6 \pi ^2}+\frac{5 \vA}{24 \pi ^2}+\frac{41 \wA}{48 \pi ^2}\right)\wAB+ \frac{45}{64 \pi ^2}\wAB^2
\nonumber \\
&+16 \newtoncoupling^2-\newtoncoupling\left(\frac{ \vAB}{4 \pi }+\frac{3  \vA}{2 \pi }+\frac{4  \wA}{3 \pi }\right)+\frac{5 \vAB^2}{64 \pi ^2}+\frac{7 \vAB \vA}{48 \pi ^2}+\frac{\vA^2}{48 \pi ^2}+\frac{\wA^2}{48 \pi ^2} \,,
\nonumber \\[4pt]
\beta_{\vA} =&\left(4-\frac{8 \newtoncoupling}{3 \pi }+\frac{5 \vAB}{48 \pi ^2}+\frac{17 \wA}{24 \pi ^2}+\frac{3 \wAB}{8 \pi ^2}\right)\vA+\frac{1}{12 \pi ^2}\vA^2
\nonumber \\
&+\newtoncoupling\left(\frac{ \wA}{2 \pi }-\frac{3 \vAB}{4 \pi }-\frac{5  \wAB}{6 \pi }\right)-\frac{\vAB^2}{96 \pi ^2}+\frac{\vAB \wAB}{96 \pi ^2}-\frac{\wA \wAB}{48 \pi ^2}-\frac{\wAB^2}{96 \pi ^2} \,,\nonumber\\[4pt]
\beta_{\vAB} =&\left(4-\frac{41 \newtoncoupling}{12 \pi }+\frac{\vA}{16 \pi ^2}+\frac{17 \wA}{24 \pi ^2}+\frac{25 \wAB}{48 \pi ^2}\right)\vAB+\frac{7}{64 \pi ^2}\vAB^2\nonumber\\
&-16 \newtoncoupling^2-\newtoncoupling\left(\frac{3  \vA}{2 \pi }+\frac{5  \wA}{3 \pi }+\frac{ \wAB}{3 \pi }\right)-\frac{\vA^2}{48 \pi ^2}+\frac{\vA \wAB}{8 \pi ^2}-\frac{\wA^2}{48 \pi ^2}-\frac{\wA \wAB}{48 \pi ^2}-\frac{\wAB^2}{64 \pi ^2}\,.
\end{align}
We first note the absence of pure gravitational inducing contributions to $\beta_{\wA}$ and $\beta_{\vA}$. This is an artefact of the chosen basis, and only occurs for $\Lambda=0$, and does therefore not have any physical interpretation. Even at $\cosmoconstant=0$, the couplings $\wA$ and $\vA$ are indirectly induced by non-vanishing fixed-point values for $\wAB$ and $\vAB$, which in turn feature direct gravitational contributions to the inducing term also at $\Lambda=0$. Away from $\cosmoconstant=0$, all these four couplings are induced by gravity, i.e., feature terms $\sim \newtoncoupling^2$ in their beta functions.
In the left panel of Fig.~\ref{fig:twogauge} we show the fixed-point structure of the two-species system for $\Lambda=-0.5$. We observe that the fixed-point values of $\wA$ and $\vA$ remain small compared to $\wAB$ and $\vAB$. This is a consequence of the small pure-gravitational inducing terms in their respective beta-functions, which cancel for $\Lambda=0$. 

As in the single-species case, the matter contributions to the scale dependences, which encode the back-reaction between both gauge fields, are suppressed  by a factor of $1/\pi$ for gravity-matter contributions, and by a factor of $1/\pi^2$ for pure matter contributions, compared to the pure-gravitational contributions, see \cite{Eichhorn:2017eht} for a detailed discussion. Therefore, as long as the fixed-point values at the sGFP remain small, the second gauge field will only result in small quantitative changes,  compared to the single-species system. 

To better quantify the impact of the second gauge field, we define the critical value of the Newton coupling $\Gcrit(\cosmoconstant)$ as the line in the $\newtoncoupling-\cosmoconstant$ plane, where the sGFP becomes complex, i.e., where the condition \eqref{eq:WGBcondition} is satisfied,
\begin{equation}
\label{eq:Gcritdef}
\Gcrit(\Lambda) = \newtoncoupling(\Lambda), \mbox{ such that } \underset{\newtoncoupling(\Lambda)}{\rm min}\,{\rm Im}[\wAast, \vAast....]\neq 0.
\end{equation}
In the right panel of Fig.~\ref{fig:twogauge} we compare the values of $\Gcrit(\cosmoconstant)$ of the single-species and two-species systems. We see that the second gauge field only changes the position of the WGB slightly, with the relative difference of $\Gcrit(\cosmoconstant)$ at the level of few percent. This is in line with the expectation that for  low numbers of species, the gravitational terms are dominant, due to the suppression by factors of $1/\pi$ of pure-matter contributions. This indicates that the two gauge fields are almost decoupled, leading to a very small impact of the second gauge field on the WGB. This observation is qualitatively similar to the WGB of induced scalar interactions, where only the inclusion of many scalar fields results in a significant change of $\Gcrit$, see \cite{deBrito:2021pyi}.
\FloatBarrier
\section{$\NV$ species of vector fields}
\label{sec:Ngauge}
We extend the previous analysis and study a coupled system of $\NV$ Abelian gauge fields. Based on our findings in the previous section, where we have seen that gravitational fluctuations respect the  global $\OT$ symmetry of the kinetic term, we will consider the generalization to  a globally $O(\NV)$ symmetric system in the following. We stress that this symmetry is realized at vanishing gauge coupling, also if the $\NV$ gauge fields transform in a local, non-Abelian symmetry group. Given that asymptotically safe gravity fluctuations preserve the asymptotically free fixed point in non-Abelian gauge couplings (at least in studies based on truncations of the full dynamics \cite{Daum:2009dn,Christiansen:2017cxa,Folkerts:2011jz}), the fixed-point structure of induced interactions of a system of $\NV=8+3+1$ gauge fields with global $O(\NV)$ symmetry is likely relevant for the Standard Model, with its 8 gluons, 3 weak gauge bosons and one Abelian gauge field. Once the gauge couplings depart from their vanishing fixed-point value (or if the Abelian gauge coupling starts out at an interacting fixed point \cite{Harst:2011zx,Eichhorn:2017lry}), additional interactions will be generated and the global $O(\NV)$ symmetry will be broken.\\
\subsection{The weak-gravity bound for many vector fields}
The most general $O(\NV)$ symmetric and gauge invariant effective action up to  dimension $8$ operators  with four gauge fields is given by  extending the range of the species index $a\in(1,\dots,\NV)$ in Eq.~\eqref{eq:gammaktwogauge}. Thus, we consider the flowing action 
\begin{align}
\label{eq:gammakON}
\Gamma_k^{U(1)^{\NV}} &= \frac{1}{4} \Int  F^{a}_{\mu\nu} F^{\mu \nu,\,a} +\,S_{\mathrm{gf},\, A}  \nonumber\\
&+ \frac{k^{-8}}{8 \NV}  \Int  \left( \w\,[F^{a}_{\mu\nu} F^{\mu\nu,\,a}]^2+\wT\,(F^a_{\mu\nu} F^{\mu\nu,\,b})(F^a_{\rho\sigma} F^{\rho\sigma,\,b})\right)   \nonumber\\
&+ \frac{k^{-8}}{8 \NV}  \Int   \left(\vt\,\left[ F^a_{\mu\nu} \tilde{F}^{\mu\nu,\,a}\right]^2+ \vT\,(F^a_{\mu\nu} \tilde{F}^{\mu\nu,\,b})(F^a_{\rho\sigma} \tilde{F}^{\rho\sigma,\,b})\right)\,,
\end{align}
where the species indices $a$ and $b$ run from 1 to $\NV$. For the choice $\NV=2$, we recover the $\OT$-symmetric system studied in Sec.~\ref{sec:twogauge}. The beta functions for $\Lambda=0$ (for the general expression see the ancillary notebook) and for general $\NV$ are given by
 \begin{align}
 \label{eq:Ngaugebetas}
 \beta_{\wA}=&\left(4-\frac{13 \newtoncoupling}{6 \pi }+\frac{5 (\NV+1) \vAB}{24 \pi ^2 \NV}+\frac{5 \vA}{12 \pi ^2 \NV}+\frac{(5 \NV+36) \wAB}{24 \pi ^2 \NV}\right)\wA+  \frac{21 \NV+13}{24 \pi ^2 \NV} \wA^2
 \nonumber\\
 &+\newtoncoupling \left(-\frac{3 \vAB}{4 \pi }+\frac{\vA}{2 \pi }-\frac{2 \wAB}{3 \pi }\right)\nonumber\\
 &+\frac{(\NV+8) \vAB^2}{192 \pi ^2 \NV}+\frac{\vAB \vA}{24 \pi ^2 \NV}+\frac{\vA^2}{8 \pi ^2 \NV}+\frac{(\NV+14) \vAB \wAB}{96 \pi ^2 \NV}+\frac{(\NV+32) \wAB^2}{192 \pi ^2 \NV}
 \,,
 \nonumber\\[4pt]
 \beta_{\wAB} =& \left(4-\frac{17 \newtoncoupling}{6 \pi }+\frac{(7 \NV+18) \vAB}{96 \pi ^2 \NV}+\frac{5 \vA}{12 \pi ^2 \NV}+\frac{(12 \NV+17) \wA}{24 \pi ^2 \NV}\right)\wAB+  \frac{23 \NV+224}{192 \pi ^2 \NV} \wAB^2
 \nonumber \\
 & +8 \newtoncoupling^2 \NV+\newtoncoupling\left(-\frac{\vAB}{4 \pi }-\frac{3 \vA}{2 \pi }-\frac{4 \wA}{3 \pi }\right)\nonumber\\
 & +\frac{(7 \NV+16) \vAB^2}{192 \pi ^2 \NV}+\frac{7 \vAB \vA}{24 \pi ^2 \NV}+\frac{\vA^2}{24 \pi ^2 \NV}+\frac{\wA^2}{24 \pi ^2 \NV}
 \,,
 \nonumber \\[4pt]
 \beta_{\vA} =&\left(4-\frac{8 \newtoncoupling}{3 \pi }+\frac{(2 \NV +1)\vAB}{24 \pi ^2 \NV}+\frac{(6 \NV+5) \wA}{12 \pi ^2 \NV}+\frac{(\NV+7) \wAB}{12 \pi ^2 \NV}\right)\vA+\frac{1}{6 \pi ^2 \NV} \vA^2
 \nonumber \\
 &\newtoncoupling\left(-\frac{3 \vAB}{4 \pi }+\frac{\wA}{2 \pi }-\frac{5 \wAB}{6 \pi }\right)-\frac{(\NV+2) \vAB^2}{192 \pi ^2 \NV}-\frac{(\NV-4) \vAB \wAB}{96 \pi ^2 \NV}-\frac{\wA \wAB}{24 \pi ^2 \NV}-\frac{(\NV+2) \wAB^2}{192 \pi ^2 \NV}
  \,,\nonumber\\[4pt]
 \beta_{\vAB} =&\left(4-\frac{41 \newtoncoupling}{12 \pi }+\frac{\vA}{8 \pi ^2 \NV}+\frac{(6 \NV+5) \wA}{12 \pi ^2 \NV}+\frac{5 (3 \NV+14) \wAB}{96 \pi ^2 \NV}\right)\vAB+\frac{5 \NV+4}{64 \pi ^2 \NV}\vAB^2
 \nonumber\\
 &-8 \newtoncoupling^2 \NV+\newtoncoupling\left(-\frac{3 \vA}{2 \pi }-\frac{5 \wA}{3 \pi }-\frac{\wAB}{3 \pi }\right)\nonumber\\
 &-\frac{\vA^2}{24 \pi ^2 \NV}+\frac{\vA \wAB}{4 \pi ^2 \NV}-\frac{\wA^2}{24 \pi ^2 \NV}-\frac{\wA \wAB}{24 \pi ^2 \NV}-\frac{(\NV+4) \wAB^2}{192 \pi ^2 \NV}\,,
 \end{align}
from which we recover the beta-functions of the $\OT$ symmetric system \eqref{eq:twogaugebetas} by choosing $\NV=2$, and the beta functions of the single-species system \eqref{eq:betaw2} and \eqref{eq:betav2} by choosing $\NV=1$, rescaling $\wA\to\wA/2$ and $\vA\to\vA/2$, identifying $\wAB=\wA/2$ and $\vAB=\vA/2$, and adding $\beta_{\wA}+\beta_{\wAB}$ and $\beta_{\vA}+\beta_{\vAB}$, respectively.
Contributions that are linear in one of the matter couplings are independent of $\NV$, while contributions that are quadratic in matter couplings feature $\NV$ independent parts, as well as parts that are suppressed by $1/\NV$. Finally, only the contributions that are independent of the matter couplings, i.e., the gravitational contribution to the inducing terms, are linear in $\NV$.
Thus, the WGB becomes stronger when increasing $\NV$, 
 because the absolute value of the coefficients $C_0$ in \eqref{eq:FPSw2} increases linearly with $\NV$, while the other coefficients remain constant to leading order. Hence, the condition \eqref{eq:WGBcondition} for the WGB is met at smaller values for the Newton couplings. Thus, at large enough $\NV$, an infinitesimally small value of the $\newtoncoupling$ is sufficient to trigger a fixed-point collision in the matter sector, i.e., we expect that $\Gcrit\to0$ as $\NV\to\infty$. 

\begin{figure}[tbp]
\centering
\includegraphics[width=.48\textwidth,clip]{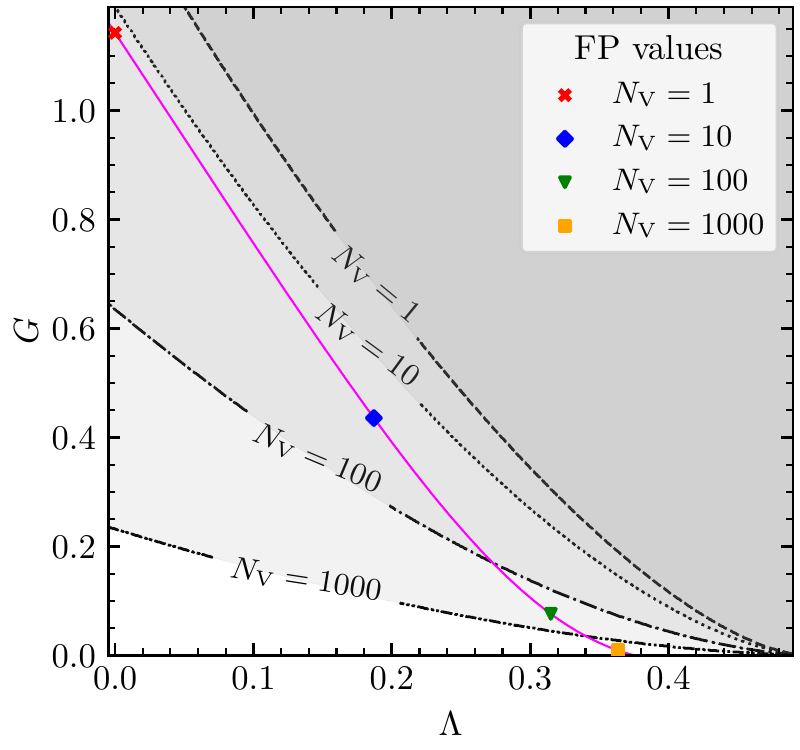}
\caption{
\label{fig:Ngauge} 
We show the WGB for the $O(\NV)$ symmetric system for different values of $\NV$, labeled by the dashed (dotted, dashdotted, dashdotdotted) line. The respective gray shaded region indicates the strong-gravity regime, which is excluded due to the presence of UV-divergences in the matter sector. The magenta line and the markers indicate the gravitational fixed-point values in the background-field approximation as a function of $\NV$, see Sec.~\ref{sec:ASandGF}.
}
\end{figure}

In Fig.~\ref{fig:Ngauge} we show the WGB for different values of $\NV$. We see that the excluded region indeed grows monotonically  with larger $\NV$. Therefore, theories with more gauge fields will have less gravitational parameter space available where the they might be UV-complete.
\subsection{Gravitational fixed-point values  in relation to the weak-gravity bound}
\label{sec:ASandGF}
So far, we have treated $\cosmoconstant$ and $\newtoncoupling$ as free input parameters. This enabled us to develop an understanding of the gravitational parameter space beyond just the behavior of the matter couplings at the gravitational fixed point.  In addition, it allows us to interpret our results in light of scenarios of effective asymptotic safety, see \cite{Percacci:2010af, deAlwis:2019aud, Held:2020kze}. In these scenarios, an effective-field theoretic description emerges from a more fundamental  theory at a cutoff scale above the Planck scale. Studying the WGB in the parameter space spanned by $\cosmoconstant$ and $\newtoncoupling$ constrains the initial conditions for $\cosmoconstant$ and $\newtoncoupling$ of this effective field theory.

Yet, it is of course crucial to supplement the analysis by the actual fixed-point values for $\newtoncoupling$ and $\cosmoconstant$ to discover whether or not asymptotic safety evades the weak-gravity bound and passes this critical viability test.  The existence of a fixed point for pure gravity has been established in numerous works \cite{Souma:1999at,Lauscher:2001ya,Reuter:2001ag,Lauscher:2002sq,Litim:2003vp,Codello:2006in,Machado:2007ea,Codello:2008vh,Benedetti:2009rx,Eichhorn:2009ah,Manrique:2010am,Eichhorn:2010tb,Groh:2010ta,Dietz:2012ic,Christiansen:2012rx,Rechenberger:2012pm,Falls:2013bv,Ohta:2013uca,Eichhorn:2013xr,Falls:2014tra,Codello:2013fpa,Christiansen:2014raa,Demmel:2015oqa,Gies:2015tca,Christiansen:2015rva,Ohta:2015fcu,Ohta:2015efa,Falls:2015qga,Eichhorn:2015bna,Gies:2016con,Denz:2016qks,Biemans:2016rvp,Falls:2016msz,Falls:2016wsa,deAlwis:2017ysy,Christiansen:2017bsy,Falls:2017lst,Houthoff:2017oam,Falls:2017cze,Becker:2017tcx,Knorr:2017fus,Knorr:2017mhu,DeBrito:2018hur,Falls:2018ylp,Bosma:2019aiu,Knorr:2019atm,Falls:2020qhj,Kluth:2020bdv,Knorr:2021slg,Bonanno:2021squ,Baldazzi:2021orb,Sen:2021ffc,Mitchell:2021qjr,Knorr:2021iwv,Baldazzi:2021fye,Fehre:2021eob} and under the inclusion of matter in \cite{Narain:2009fy,Dona:2012am,Percacci:2015wwa, Labus:2015ska, Christiansen:2017cxa, Dona:2013qba,Meibohm:2015twa,Dona:2015tnf,Biemans:2017zca,Alkofer:2018fxj,Wetterich:2019zdo,Sen:2021ffc, Eichhorn:2018akn, Eichhorn:2018ydy,Eichhorn:2018nda, Burger:2019upn, Daas:2020dyo, Eichhorn:2020sbo, Daas:2021abx, Laporte:2021kyp}.
To search for fixed points, we use the beta-functions for $\cosmoconstant$ and $\newtoncoupling$ \cite{Eichhorn:2016vvy}, supplemented with the matter contributions from \cite{Dona:2012am, Dona:2013qba}.
We 
find that the fixed-point values stay below the WGB for all values of $\NV$ that we investigate explicitly. In particular, this means that systems with 12 gauge fields, like the SM, evade the corresponding weak-gravity bound. 

Our numerical investigation stops at a finite, large $\NV$, and we supplement it by an analytical large $\NV$ study:
To determine whether or not the WGB is still evaded for $\NV\to\infty$, we evaluate the beta functions of the induced matter couplings \eqref{eq:Ngaugebetas} at the gravitational fixed-point values for $\newtoncoupling$ and $\cosmoconstant$. The resulting beta functions feature a well-defined $\NV\to\infty$ limit, given by
\begin{align}
\label{eq:NVlimit}
\beta_{\wA}&= 
\left(\frac{5 \vAB}{24 \pi ^2}+\frac{5 \wAB}{24 \pi ^2}+4 \right)\wA + \frac{7}{8 \pi ^2}
 \wA^2 +\left( \frac{\vAB^2}{192 \pi ^2}+\frac{\vAB \wAB}{96 \pi ^2}+\frac{\wAB^2}{192 \pi ^2} \right)\,,
\nonumber\\
\beta_{\wAB}&= 
\left( \frac{7 \vAB}{96 \pi ^2}+\frac{\wA}{2 \pi ^2}+4 \right)\wAB + \frac{23}{192 \pi ^2}
\wAB^2 +  \frac{7 \vAB^2}{192 \pi ^2} \,,
\nonumber\\
\beta_{\vA}&= 
\left( \frac{\vAB}{12 \pi ^2}+\frac{\wA}{2 \pi ^2}+\frac{\wAB}{12 \pi ^2}+4 \right)\vA +\left(  -\frac{\vAB^2}{192 \pi ^2}-\frac{\vAB \wAB}{96 \pi ^2}-\frac{\wAB^2}{192 \pi ^2} \right)\,,
\nonumber\\
\beta_{\vAB}&= 
\left( \frac{\wA}{2 \pi ^2}+\frac{5 \wAB}{32 \pi ^2}+4 \right)\vAB 
+\frac{5}{64 \pi ^2}\vAB^2 -\frac{\wAB^2}{192 \pi ^2} \,.
\end{align}
The existence of this limit indicates that ${\newtoncoupling}_*\to0$ faster than $1/\sqrt{\NV}$ as $\NV$ increases, cf.~\eqref{eq:Ngaugebetas}. Hence, gravity and matter become weakly coupled for large $\NV$ and decouple entirely in the $\NV\to\infty$ limit. Since gravity and matter are decoupled for $\NV\to\infty$, the self-interactions are no longer induced in this limit, but assume their Gaussian fixed-point values.

In summary, our studies indicate that asymptotically safe quantum gravity  could be compatible with any number of $O(\NV)$-symmetric gauge fields. Therefore, within our truncation and the investigated mechanism, the interplay of gravity and matter does not give rise to any bound on the number of vector fields. This is in contrast to the scalar sector, where no number of scalar fields (in the absence of spinning matter) gives rise to a UV-complete scalar-gravity system \cite{deBrito:2021pyi}.

\section{Gauge dependence as a test of the robustness of the results}
\label{sec:gaugedependence}
In our discussion so far, we have fixed the gauge parameter $\betagauge=1$. In the following, we will investigate the robustness of the WGB under variations of $\betagauge\in[-\infty,3)$ in the Landau limit, i.e., $\alpha_A \to 0$ and $\alpha_h \to 0$. The point $\betagauge =3$ is an incomplete gauge fixing, thus results for values $\betagauge \lesssim 3$ become untrustworthy.

This serves as a test for the robustness of our truncation:  physical information, such as the existence of a fixed point, and physical quantities, such as the critical exponents, should be independent of the gauge parameters, when $\Gamma_k$ is not truncated. Conversely, gauge dependence arises in physical quantities, when truncations are employed.
When extending the truncation, these gauge dependences are expected to decrease, converging to gauge-independent physical quantities when $\Gamma_k$ is not truncated at all. Similarly, when a truncation already captures the physics of a system well, gauge dependences are expected to be mild. In this spirit, we explore the gauge dependence of our results as a test of their robustness.

In the following, we  focus on the WGB for the different systems discussed in the previous sections, namely a single species, two species and $N_V$ species of gauge fields.   For simplicity, we specify to $\cosmoconstant=0$, and compute the critical value $\Gcrit(\cosmoconstant=0)$ as a function of $\betagauge$, cf.~Eq.~\eqref{eq:Gcritdef}. 

By investigating $\Gcrit$ for the specific choice $\cosmoconstant=0$ as a function of $\betagauge$, we gain insight into the robustness of the WGB.
\subsection{One species of gauge fields}
We follow a similar strategy as in the previous sections, and first investigate the gauge-dependence of the WGB for two truncations, which contain either only $\wA$ or only $\vA$  and consider a third truncation which contains both couplings as a second step.
 
\begin{figure}[tbp]
	\centering
	\includegraphics[width=.5\textwidth]{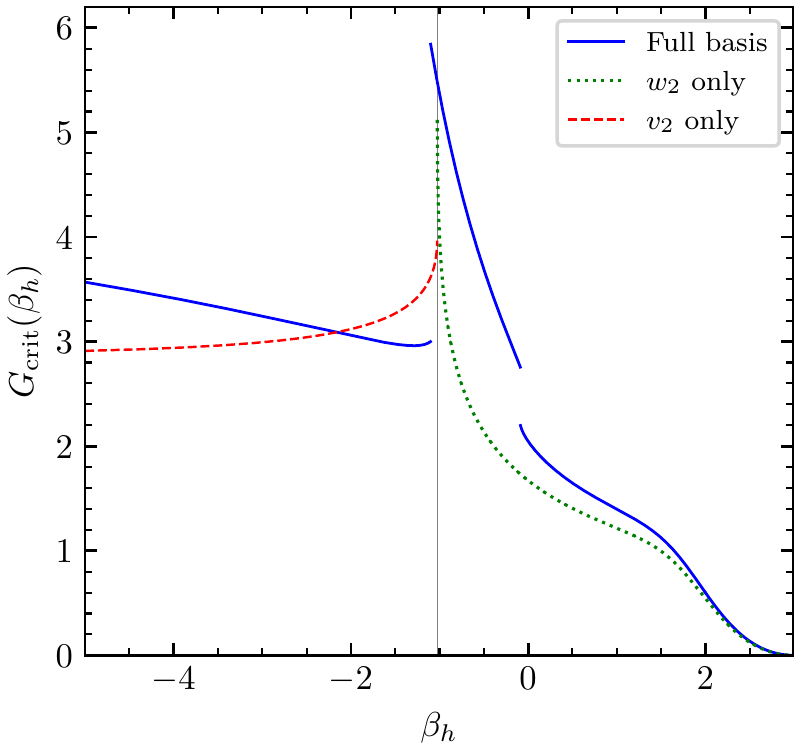}
	\caption{We show the critical value of the Newton coupling $\Gcrit$, where the sGFP becomes complex, for $\Lambda=0$, and as a function of the gauge-fixing parameter $\betagauge$. The green (dotted) line shows $\Gcrit$ in a truncation only containing the induced interaction $\wA$, where the WGB is only present for $\betagauge\gtrsim-1.03$ (indicated by the vertical gray line). The red (dashed) line shows $\Gcrit$ in a truncation only containing the induced interaction $\vA$, where the WGB is only present for $\betagauge\lesssim-1.03$. The blue (solid) line shows $\Gcrit$ in the full set of induced four-gauge interactions at mass dimension $8$. In the full system, the WGB exists for all values of $\betagauge$.
	}
	\label{fig:Gcrit} 
\end{figure}

For the individual couplings, the relative sign between the coefficients $C_{0}$ and $C_{2}$ determines if a WGB is possible in principle, cf.~Eq.~\eqref{eq:FPSw2}. Since only pure gauge-field  diagrams contribute to $C_{2}$,  it is independent of $\betagauge$.\\
For a truncation with $\wA$ only, the gauge-dependent coefficient $C_{0,\wA}(\newtoncoupling, \vA=0)$ reads
\begin{align}
C_{0,\wA}(\newtoncoupling, \vA=0) = \frac{4 \newtoncoupling^2}{3 (-3 +\beta)^4} \left(225-60\beta-154 \beta^2 + 100 \beta^3 - 15 \beta^4\right)\,,
\end{align}
which is positive for $\betagauge\gtrsim-1.03$, and negative for smaller $\betagauge$. Thus, for $\betagauge \lesssim -1.03$, there is no WGB in a truncation with only $\wA$. For larger $\betagauge$, a WGB exists, and the resulting $\Gcrit$ at $\Lambda=0$ only quantitatively depends on $\betagauge$, see Fig.~\ref{fig:Gcrit}.

The situation is reversed when considering $\vA$ as the only interaction of the system, since
\begin{equation}
C_{0,\vA}(\newtoncoupling, \wA=0) = - C_{0,\wA}(\newtoncoupling, \vA=0)\,.
\end{equation}
As $C_{2, \vA}>0$, cf.~Eq.~\eqref{eq:betav2}, a WGB for $\vA$ can only exist for $\betagauge\lesssim-1.03$, cf.~Fig.~\ref{fig:Gcrit}.

We have seen in Sec.~\ref{sec:onegauge} that the  effect of $\vA$ in $\beta_{\wA}$ and vice-versa is suppressed compared to the gravitational contributions. Therefore,  we expect that the  truncation containing both  $\wA$ and $\vA$ features a WGB that is quantitatively close to the sum of the two individual WGBs for all values of $\betagauge$. An explicit computation of $\Gcrit$ confirms this expectation, and $\Gcrit$ is finite and non-zero for all $\betagauge\in[-\infty,3)$, see Fig.~\ref{fig:Gcrit}. This result indicates the reliability of the previous results \cite{Christiansen:2017gtg, Eichhorn:2019yzm} for the choice $\betagauge=1$ that was employed in these works, but also highlights the importance of considering a full basis of induced operators at a given mass dimension, to reliably investigate the presence of a WGB.
	
In summary, the WGB of the full single gauge-field system features a WGB for all values of the gauge parameter $\betagauge
<3$. The position of the WGB in the gravitational parameter space only depends quantitatively on $\betagauge$. This indicates that indeed the physical information, namely that a strongly coupled regime of quantum gravity appears to be incompatible with a UV-complete gauge sector, is independent of the gauge choice.

\subsection{Multiple species of gauge fields}
For an estimate on the gauge-dependence of the $\OT$-symmetric system discussed in Sec.~\ref{sec:twogauge}, we compare the location of the WGB with the single-species system for various choices of $\betagauge$, see the right panel of Fig.~\ref{fig:twogauge}. For each displayed gauge-choice, the difference of the WGB in the single-species and two-species systems is at the level of a few percent. An exception to this small deviation appears for $\Lambda\approx0$, which is produced by a sign-flip in the beta-functions, giving rise to a dent in the WGB. We comment on this feature in App.~\ref{sec:WGBbetazero}. The overall small deviation of the WGB between the single-species and two-species systems indicates that the impact of the second gauge field on the WGB remains small, independent of the gauge choice and that the gauge dependence is overall not very large, except close to $\betagauge =-1$, where somewhat larger variations occur.

\begin{figure}[tbp]
\includegraphics[width=.475\textwidth,clip]{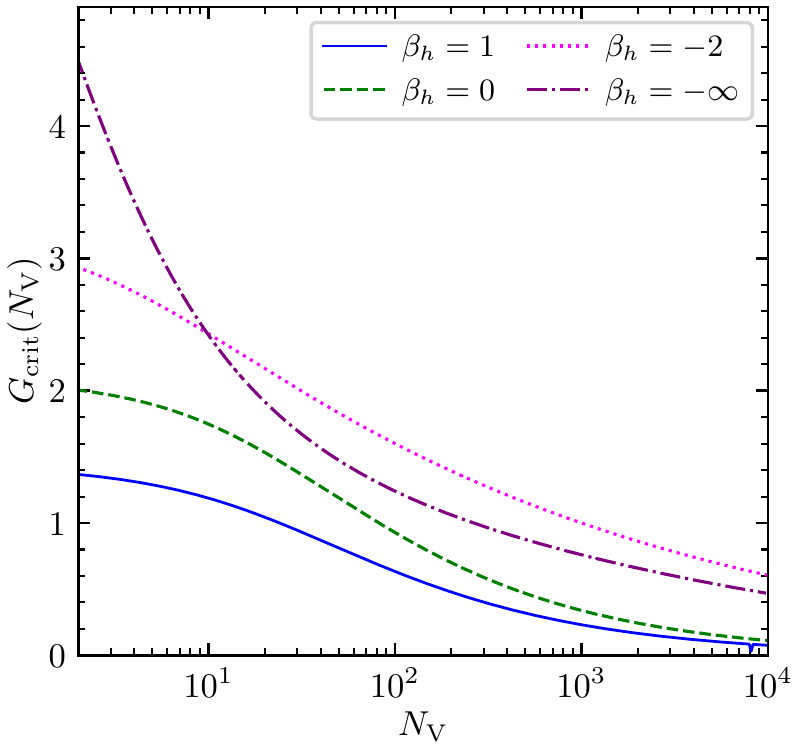}
\hfill
\includegraphics[width=.475\textwidth,clip]{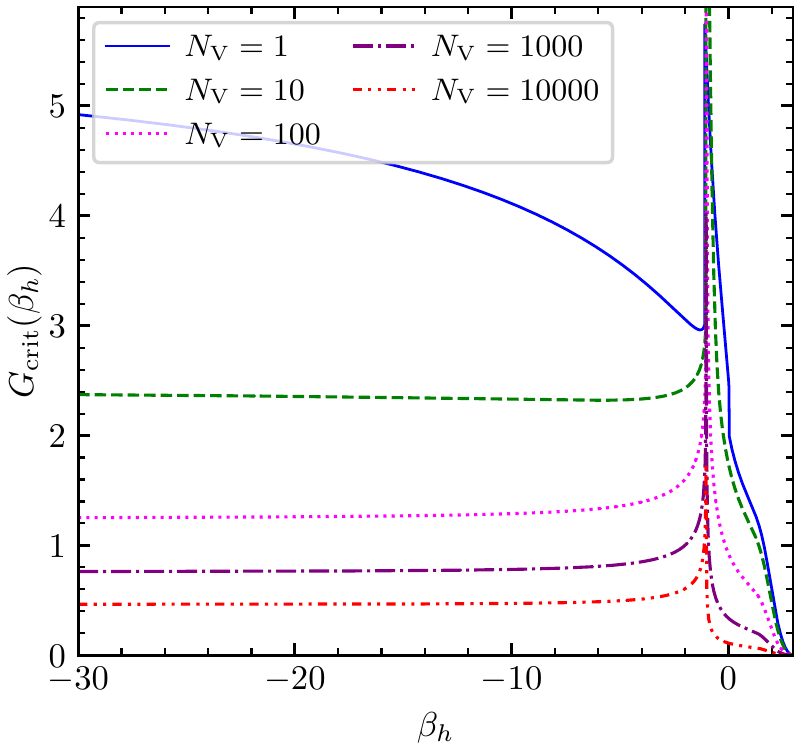}
\caption{
\label{fig:manygaugebetadependence} 
We show the WGB in a function of $\NV$ and $\betagauge$. $G^{\mathrm{crit}}$ denotes the value of $\newtoncoupling^{\ast}$ at which the sGFP becomes complex. \newline
To the left: Dependence of the $\Gcrit(\cosmoconstant=0)$ on $\NV$ given selected values of $\betagauge$.\newline
To the right: Dependence of the $\Gcrit(\cosmoconstant=0)$ on $\betagauge$ given selected values of $\NV$.}
\end{figure}

We study the gauge-dependence of the $O(\NV)$-symmetric system by computing the critical value $\Gcrit$ at $\Lambda=0$. As in the single- and two-species systems, $\Gcrit$ depends on $\betagauge$ only quantitatively, and remains finite for all displayed values of $\NV$ and $\betagauge$, see Fig.~\ref{fig:manygaugebetadependence}. The strongest dependence on $\betagauge$ is found for $-5\lesssim\betagauge<3$, see the right panel of Fig.~\ref{fig:manygaugebetadependence}. This is a consequence of the incompleteness of the gauge fixing at $\betagauge=3$. Since these gravitational contributions at $\Lambda=0$ are functions in $(\betagauge-3)^{-n}$, where $n\in\mathbb{Z}_+$, their variation with $\betagauge$ decreases for more negative $\betagauge$. 

Furthermore, the overall gauge dependence decreases when increasing the number of gauge fields.
This is because the contribution from the gauge fields is independent of $\betagauge$, and starts to become more important than the gravitational contribution, once $N_V$ is large enough.  At these values of $N_V$, the WGB is only weakly dependent on $\betagauge$ \footnote{This is easier to see after rescaling the matter couplings with $\NV$. This rescaling is just a redefinition of the couplings and does not change the location of the WGB. In the rescaled version of the beta functions, the matter contributions is proportional to $\NV$, while the gravitational contribution is $\NV$-independent.}.
Therefore, the qualitative feature that the inclusion of more gauge fields result in a stronger WGB, is independent of the gauge choice. 

\section{Conclusions and discussion}
\label{sec:discussion}
In this work, we have developed a deeper and more extensive understanding of the weak-gravity bound, discovered in \cite{Christiansen:2017gtg} for gauge-gravity systems.\\
First, we have extended previous work \cite{Christiansen:2017gtg,Eichhorn:2019yzm} by the second 4-photon-interaction, namely $(F \tilde{F})^2$. Our first key result shows the robustness of previous studies: Including the additional term changes the critical strength of gravity, i.e., the weak-gravity bound, only very slightly. This nontrivial result could even be interpreted as an indication for the onset of apparent convergence of the weak-gravity bound under extensions of the truncation of the dynamics.\\
Further we strengthen the evidence, first found in \cite{Eichhorn:2019yzm}, that spacetime dimensionalities close to four are preferred by asymptotically safe gravity-matter systems. The mechanism is simple: To render the Abelian gauge coupling asymptotically free or safe, and solve its triviality problem, the strength of metric fluctuations must increase as a function of dimensionality beyond four. At the same time, the weak-gravity bound continues to exist in dimensionality beyond four, and prohibit gravitational fixed-point values to enter the regime where they are strong enough to solve the Abelian triviality problem.\\
We also consider systems with more than one gauge field for the first time. Here, we observe results similar to scalar-gravity systems \cite{deBrito:2021pyi}: First, the weak-gravity bound becomes stronger with increasing number of fields. Second, the increase is slow, i.e., the weak-gravity bound at 12 gauge fields (as in the SM) is still quantitatively close to the weak-gravity bound at one gauge field. Unlike for scalar-gravity systems, the gravitational fixed-point values lie below the weak-gravity bound for any number of gauge fields. The latter statement is subject to systematic uncertainties, both on the location of the weak gravity bound (which is not very large, if the change under the inclusion of $(F \tilde{F})^2$ provides a robust estimate) and on the location of the gravitational fixed-point values.\\
Our fourth key result  shows the robustness of the weak-gravity bound: We use gauge dependence as a proxy for systematic uncertainties; because physical results (such as the existence of a bound, beyond which no asymptotically safe fixed point can exist) exhibit gauge dependence when truncations of the dynamics are used, the amount of gauge dependence quantifies the impact of terms beyond the truncation. We find that the weak-gravity bound is stable under variations of a gravitational gauge parameter. To obtain this result, the inclusion of the second type of four-photon interaction, $(F \tilde{F})^2$, is crucial: for a range of values of the gauge parameter, the mechanism behind the weak-gravity bound is exhibited by $(F^2)^2$; for the neighboring range, it is exhibited by $(F \tilde{F})^2$. Thus, our result shows not just the stability of the weak-gravity bound, but also the importance of including a full basis of interactions at the interaction-order of interest.\\
Our results are subject to systematic uncertainties from our choice of truncation, as discussed above. Further, we have used the Functional Renormalization Group in Euclidean spacetime to obtain our results. Therefore, one may well wonder about their applicability to Lorentzian spacetime. In this context, the weak-gravity bound could turn out to be an important piece of information: While an analytical continuation from Euclidean to Lorentzian signature is not possible in full quantum gravity, or even on a general background, it is (under certain conditions on the propagators, currently under investigation, e.g., in \cite{Bosma:2019aiu,Platania:2020knd,Bonanno:2021squ,Fehre:2021eob}), possible around a perturbative, flat background. One might interpret the weak-gravity bound as an indication that asymptotically safe gravity dynamically prefers this more weakly-coupled regime, in which fluctuations about a flat background might be a good approximation to the dynamics of full quantum gravity. In turn, this would make an analytical continuation potentially feasible, and our results therefore relevant for Lorentzian signature.


\acknowledgments
We would like to thank Benjamin Knorr, Gustavo de Brito and Rafael Robson Lino dos Santos for insightful discussions.
A.~E.~is supported by a research grant (29405) from VILLUM FONDEN and
J.~H.~K.~acknowledges the NAWA Iwanowska scholarship PPN/IWA/2019/1/00048.  The research of M.~S.~has been supported by a scholarship of the German Academic Scholarship Foundation and by the Perimeter Institute for Theoretical Physics. Research at Perimeter Institute is supported in part by the Government of Canada through the Department of Innovation, Science and Economic Development and by the Province of Ontario through the Ministry of Colleges and Universities.  M.~S.~ and J.~H.~K.~are grateful to CP3-Origins at the University of Southern Denmark for extended hospitality during various stages of this work.

\appendix
\section{Cases in which the Weak Gravity Bound is not a function in the $G-\Lambda$ plane}
\label{sec:WGBbetazero}
\begin{figure}[tbp]
\includegraphics[width=.475\textwidth,clip]{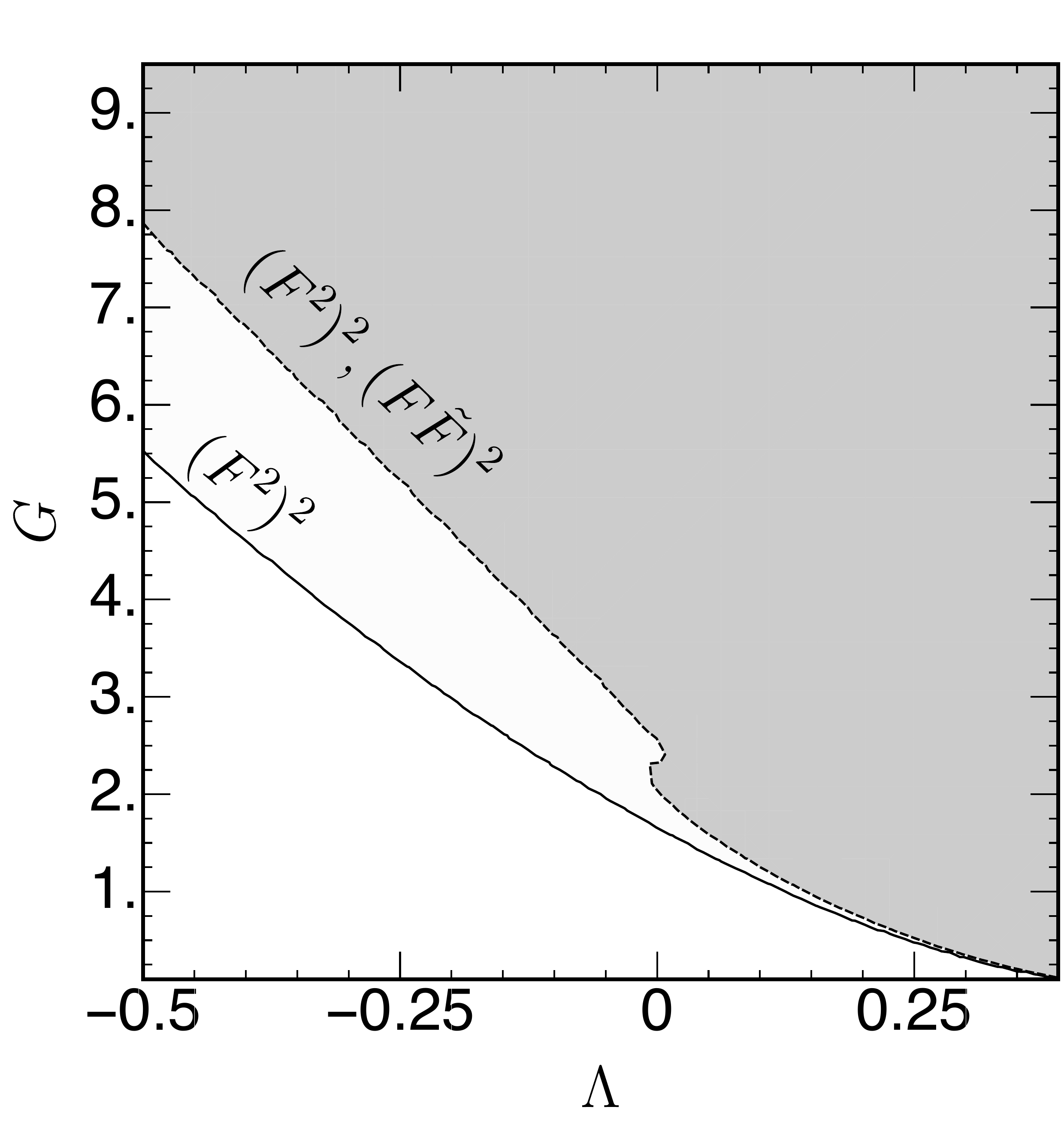}
\hfill
\includegraphics[width=.475\textwidth,clip]{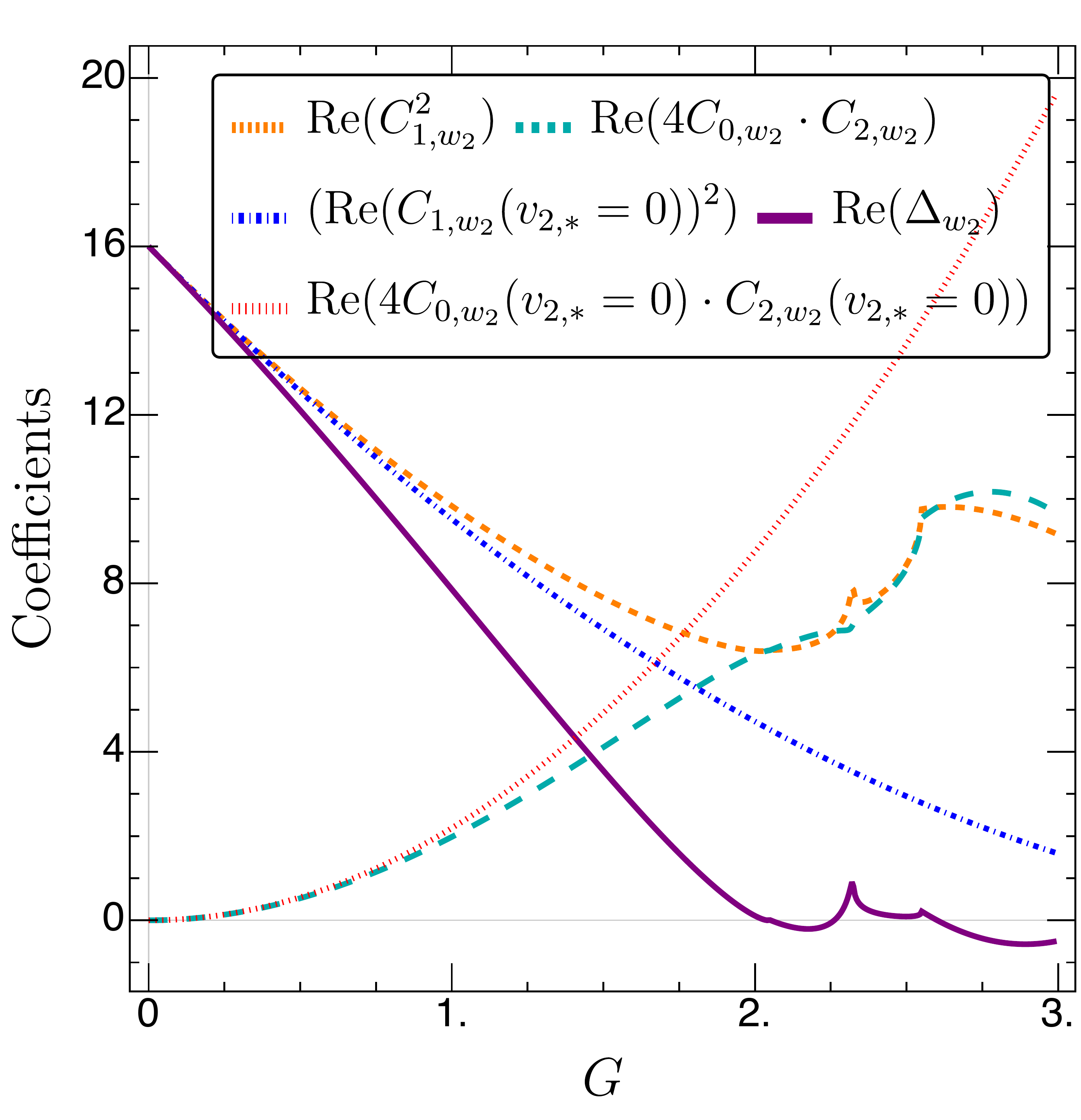}
\caption{
\label{fig:onegaugebetadependence} 
In the left panel: Comparison of the WGB for $\wA$ and $\wA$ plus $\vA$ truncations for $\betagauge=0$. For the full truncation there is a kink at $\Lambda \approx 0$ absent for $\wA$ truncation only.\newline
In the right panel: The behavior of coefficients of the coupled $\wA$ and $\vA$ system. 
}
\end{figure}

The $\betagauge\approx0$ case deserves special attention, since the WGB features a dent in the $\newtoncoupling-\Lambda$ plane, see Fig.~\ref{fig:onegaugebetadependence}, which leads to a discontinuity in $\Gcrit$, cf.~Fig~\ref{fig:Gcrit}. Here we discuss it for the representative case of the single species system for \betagauge=0. The WGB for $\betagauge=0$ is not a function for $\cosmoconstant\approx0$, see the left panel of the Fig.~\ref{fig:onegaugebetadependence}. This is different for other gauge choices, where $G_{\textrm{N}, \, \rm crit}(\Lambda)$ is a function, i.e., there is a single value $G_{\textrm{N}, \, \rm crit}$ for a given value of $\Lambda$.
This behavior is present only for the full truncation consisting of $\wA$ and $\vA$ and absent for truncations consisting of $\wA$ or $\vA$ only. \\
In particular, for $\cosmoconstant=0$ there are two allowed intervals, $\newtoncoupling <2.05$ and $ 2.31<\newtoncoupling <2.55$. This follows from the form of beta function for $\wA$. Let us consider the schematic form of $\beta_{\wA}$, which we repeat here for convenience
 \begin{equation}
 \beta_{\wA}=  C_{2,\wA}(\newtoncoupling,\vA,\cosmoconstant=0) \wA^2+C_{1,\wA} (\newtoncoupling, \vA, \cosmoconstant=0) \wA +C_{0,\wA} (\newtoncoupling,\vA,\cosmoconstant=0).
 \end{equation}
 As long as the discriminant $\Delta = C_{1,\wA}^2 - 4C_{0,\wA}C_{2,\wA}$ is non-negative, the real shifted Gaussian FP exist. At $\Delta = 0$, there is collision of the fixed points. For $\vA\equiv0$ the $\textrm{Re}(C_{1,\wA}^2(\vA=0))$ decreases with $\newtoncoupling$ (the blue dot-dashed line in right panel of Fig.~\ref{fig:onegaugebetadependence}), while $\textrm{Re}(4 C_{0,\wA} C_{2,\wA}(\vA=0))$ increases (the red dotted line). As they cross at $\newtoncoupling \approx 1.75$ the $\textrm{Re}(\Delta_{\wA})$ becomes negative and the fixed points are no-longer real. On the other hand, the contribution from $\vA$ to the $C_1$ have opposite sign to the one from $\newtoncoupling$
  \begin{align}
 \beta_{\wA}=  \left(4-\frac{155 \newtoncoupling}{54\pi}+ \frac{\vA}{12\pi^2}\right)\wA + \left(\frac{100 \newtoncoupling^2}{27}-\frac{\newtoncoupling \vA}{\pi} + \frac{\vA^2}{6\pi^2}\right)+ \frac{35}{24\pi^2} \wA^2.
 \end{align}
Since $v_{2,*}>0$, the $\textrm{Re}(C_{1,\wA}^2)$ starts to grow at $\newtoncoupling \approx 2.0$ (in the right panel of Fig.~\ref{fig:onegaugebetadependence} depicted as the dashed orange line). Furthermore the contributions from $v_{2,*}$ to the $\textrm{Re}(C_{0,\wA})$ are such that $\textrm{Re}(C_{0,\wA})<\textrm{Re}(C_{0,\wA}(v_{2,*}=0))$, depicted as the cyan dashed line. This two effects combined results in $\textrm{Re}(\Delta_{\wA})$ (the purple line) being slightly positive in the interval $ 2.31<\newtoncoupling <2.55$, resulting in a second allowed interval. This behavior is a reminder that one has to be cautious when studying systems with multiple couplings. 

The presence of a second allowed interval for $\newtoncoupling$ at fixed $\Lambda$ gives rise to a jump in $\Gcrit(\Lambda)$, cf.~the left panel of Fig.~\ref{fig:onegaugebetadependence}. The exact value for $\Lambda$ where this jump happens depends on the gauge parameter $\betagauge$, and on the number of vector fields $\NV$. Therefore, the discussed property of the WGB leads to the jump in $\Gcrit(\beta_h)$ for the single-species system, cf.~the blue (solid) line in Fig.~\ref{fig:Gcrit}. It also leads to the seemingly large difference between the WGB of the single-species and two-species system around $\Lambda=0$, cf.~the right panel of Fig.~\ref{fig:twogauge}, since the jump in $\Gcrit$ occurs for slightly different values of $\Lambda$ in both systems.
\FloatBarrier

\addcontentsline{toc}{section}{The Bibliography}
\bibliography{references.bib,manuals.bib}{}

\providecommand{\href}[2]{#2}\begingroup\raggedright\begin{thebibliography}{100}

\bibitem{Addazi:2021xuf}
A.~Addazi et~al., \emph{{Quantum gravity phenomenology at the dawn of the
  multi-messenger era -- A review}},
  \href{https://arxiv.org/abs/2111.05659}{{\ttfamily 2111.05659}}.

\bibitem{Eichhorn:2017egq}
A.~Eichhorn, \emph{{Status of the asymptotic safety paradigm for quantum
  gravity and matter}},
  \href{https://doi.org/10.1007/s10701-018-0196-6}{\emph{Found. Phys.}
  {\bfseries 48} (2018) 1407}
  [\href{https://arxiv.org/abs/1709.03696}{{\ttfamily 1709.03696}}].

\bibitem{Percacci:2017fkn}
R.~Percacci, \emph{{An Introduction to Covariant Quantum Gravity and Asymptotic
  Safety}}, vol.~3 of \emph{100 Years of General Relativity}, World Scientific
  (2017), \href{https://doi.org/10.1142/10369}{10.1142/10369}.

\bibitem{Eichhorn:2018yfc}
A.~Eichhorn, \emph{{An asymptotically safe guide to quantum gravity and
  matter}}, \href{https://doi.org/10.3389/fspas.2018.00047}{\emph{Front.
  Astron. Space Sci.} {\bfseries 5} (2019) 47}
  [\href{https://arxiv.org/abs/1810.07615}{{\ttfamily 1810.07615}}].

\bibitem{Reuter:2019byg}
M.~Reuter and F.~Saueressig, \emph{{Quantum Gravity and the Functional
  Renormalization Group}: {The Road towards Asymptotic Safety}}, Cambridge
  University Press (1, 2019).

\bibitem{Pereira:2019dbn}
A.D.~Pereira, \emph{{Quantum spacetime and the renormalization group: Progress
  and visions}},  in \emph{{Progress and Visions in Quantum Theory in View of
  Gravity}: {Bridging foundations of physics and mathematics}}, 4, 2019
  [\href{https://arxiv.org/abs/1904.07042}{{\ttfamily 1904.07042}}].

\bibitem{Reichert:2020mja}
M.~Reichert, \emph{{Lecture notes: Functional Renormalisation Group and
  Asymptotically Safe Quantum Gravity}},
  \href{https://doi.org/10.22323/1.384.0005}{\emph{PoS} {\bfseries 384} (2020)
  005}.

\bibitem{Pawlowski:2020qer}
J.M.~Pawlowski and M.~Reichert, \emph{{Quantum Gravity: A Fluctuating Point of
  View}}, \href{https://doi.org/10.3389/fphy.2020.551848}{\emph{Front. in
  Phys.} {\bfseries 8} (2021) 551848}
  [\href{https://arxiv.org/abs/2007.10353}{{\ttfamily 2007.10353}}].

\bibitem{Bonanno:2020bil}
A.~Bonanno, A.~Eichhorn, H.~Gies, J.M.~Pawlowski, R.~Percacci, M.~Reuter
  et~al., \emph{{Critical reflections on asymptotically safe gravity}},
  \href{https://doi.org/10.3389/fphy.2020.00269}{\emph{Front. in Phys.}
  {\bfseries 8} (2020) 269} [\href{https://arxiv.org/abs/2004.06810}{{\ttfamily
  2004.06810}}].

\bibitem{Eichhorn:2016esv}
A.~Eichhorn, A.~Held and J.M.~Pawlowski, \emph{{Quantum-gravity effects on a
  Higgs-Yukawa model}},
  \href{https://doi.org/10.1103/PhysRevD.94.104027}{\emph{Phys. Rev. D}
  {\bfseries 94} (2016) 104027}
  [\href{https://arxiv.org/abs/1604.02041}{{\ttfamily 1604.02041}}].

\bibitem{Eichhorn:2017eht}
A.~Eichhorn and A.~Held, \emph{{Viability of quantum-gravity induced
  ultraviolet completions for matter}},
  \href{https://doi.org/10.1103/PhysRevD.96.086025}{\emph{Phys. Rev. D}
  {\bfseries 96} (2017) 086025}
  [\href{https://arxiv.org/abs/1705.02342}{{\ttfamily 1705.02342}}].

\bibitem{Eichhorn:2017ylw}
A.~Eichhorn and A.~Held, \emph{{Top mass from asymptotic safety}},
  \href{https://doi.org/10.1016/j.physletb.2017.12.040}{\emph{Phys. Lett. B}
  {\bfseries 777} (2018) 217}
  [\href{https://arxiv.org/abs/1707.01107}{{\ttfamily 1707.01107}}].

\bibitem{DeBrito:2019rrh}
G.P.~De~Brito, Y.~Hamada, A.D.~Pereira and M.~Yamada, \emph{{On the impact of
  Majorana masses in gravity-matter systems}},
  \href{https://doi.org/10.1007/JHEP08(2019)142}{\emph{JHEP} {\bfseries 08}
  (2019) 142} [\href{https://arxiv.org/abs/1905.11114}{{\ttfamily
  1905.11114}}].

\bibitem{Eichhorn:2020sbo}
A.~Eichhorn and M.~Pauly, \emph{{Constraining power of asymptotic safety for
  scalar fields}},
  \href{https://doi.org/10.1103/PhysRevD.103.026006}{\emph{Phys. Rev. D}
  {\bfseries 103} (2021) 026006}
  [\href{https://arxiv.org/abs/2009.13543}{{\ttfamily 2009.13543}}].

\bibitem{Eichhorn:2017als}
A.~Eichhorn, Y.~Hamada, J.~Lumma and M.~Yamada, \emph{{Quantum gravity
  fluctuations flatten the Planck-scale Higgs potential}},
  \href{https://doi.org/10.1103/PhysRevD.97.086004}{\emph{Phys. Rev. D}
  {\bfseries 97} (2018) 086004}
  [\href{https://arxiv.org/abs/1712.00319}{{\ttfamily 1712.00319}}].

\bibitem{Grabowski:2018fjj}
F.~Grabowski, J.H.~Kwapisz and K.A.~Meissner, \emph{{Asymptotic safety and
  Conformal Standard Model}},
  \href{https://doi.org/10.1103/PhysRevD.99.115029}{\emph{Phys. Rev. D}
  {\bfseries 99} (2019) 115029}
  [\href{https://arxiv.org/abs/1810.08461}{{\ttfamily 1810.08461}}].

\bibitem{Eichhorn:2019dhg}
A.~Eichhorn, A.~Held and C.~Wetterich, \emph{{Predictive power of grand
  unification from quantum gravity}},
  \href{https://doi.org/10.1007/JHEP08(2020)111}{\emph{JHEP} {\bfseries 08}
  (2020) 111} [\href{https://arxiv.org/abs/1909.07318}{{\ttfamily
  1909.07318}}].

\bibitem{Kwapisz:2019wrl}
J.H.~Kwapisz, \emph{{Asymptotic safety, the Higgs boson mass, and beyond the
  standard model physics}},
  \href{https://doi.org/10.1103/PhysRevD.100.115001}{\emph{Phys. Rev. D}
  {\bfseries 100} (2019) 115001}
  [\href{https://arxiv.org/abs/1907.12521}{{\ttfamily 1907.12521}}].

\bibitem{Reichert:2019car}
M.~Reichert and J.~Smirnov, \emph{{Dark Matter meets Quantum Gravity}},
  \href{https://doi.org/10.1103/PhysRevD.101.063015}{\emph{Phys. Rev. D}
  {\bfseries 101} (2020) 063015}
  [\href{https://arxiv.org/abs/1911.00012}{{\ttfamily 1911.00012}}].

\bibitem{Hamada:2020vnf}
Y.~Hamada, K.~Tsumura and M.~Yamada, \emph{{Scalegenesis and fermionic dark
  matters in the flatland scenario}},
  \href{https://doi.org/10.1140/epjc/s10052-020-7929-3}{\emph{Eur. Phys. J. C}
  {\bfseries 80} (2020) 368}
  [\href{https://arxiv.org/abs/2002.03666}{{\ttfamily 2002.03666}}].

\bibitem{Kowalska:2020gie}
K.~Kowalska, E.M.~Sessolo and Y.~Yamamoto, \emph{{Flavor anomalies from
  asymptotically safe gravity}},
  \href{https://doi.org/10.1140/epjc/s10052-021-09072-1}{\emph{Eur. Phys. J. C}
  {\bfseries 81} (2021) 272}
  [\href{https://arxiv.org/abs/2007.03567}{{\ttfamily 2007.03567}}].

\bibitem{Eichhorn:2020kca}
A.~Eichhorn and M.~Pauly, \emph{{Safety in darkness: Higgs portal to simple
  Yukawa systems}},
  \href{https://doi.org/10.1016/j.physletb.2021.136455}{\emph{Phys. Lett. B}
  {\bfseries 819} (2021) 136455}
  [\href{https://arxiv.org/abs/2005.03661}{{\ttfamily 2005.03661}}].

\bibitem{Kowalska:2020zve}
K.~Kowalska and E.M.~Sessolo, \emph{{Minimal models for g-2 and dark matter
  confront asymptotic safety}},
  \href{https://doi.org/10.1103/PhysRevD.103.115032}{\emph{Phys. Rev. D}
  {\bfseries 103} (2021) 115032}
  [\href{https://arxiv.org/abs/2012.15200}{{\ttfamily 2012.15200}}].

\bibitem{Eichhorn:2012va}
A.~Eichhorn, \emph{{Quantum-gravity-induced matter self-interactions in the
  asymptotic-safety scenario}},
  \href{https://doi.org/10.1103/PhysRevD.86.105021}{\emph{Phys. Rev. D}
  {\bfseries 86} (2012) 105021}
  [\href{https://arxiv.org/abs/1204.0965}{{\ttfamily 1204.0965}}].

\bibitem{Christiansen:2017gtg}
N.~Christiansen and A.~Eichhorn, \emph{{An asymptotically safe solution to the
  U(1) triviality problem}},
  \href{https://doi.org/10.1016/j.physletb.2017.04.047}{\emph{Phys. Lett. B}
  {\bfseries 770} (2017) 154}
  [\href{https://arxiv.org/abs/1702.07724}{{\ttfamily 1702.07724}}].

\bibitem{Eichhorn:2019yzm}
A.~Eichhorn and M.~Schiffer, \emph{{$d=4$ as the critical dimensionality of
  asymptotically safe interactions}},
  \href{https://doi.org/10.1016/j.physletb.2019.05.005}{\emph{Phys. Lett. B}
  {\bfseries 793} (2019) 383}
  [\href{https://arxiv.org/abs/1902.06479}{{\ttfamily 1902.06479}}].

\bibitem{deBrito:2021pyi}
G.P.~de~Brito, A.~Eichhorn and R.R.L.d.~Santos, \emph{{The weak-gravity bound
  and the need for spin in asymptotically safe matter-gravity models}},
  \href{https://doi.org/10.1007/JHEP11(2021)110}{\emph{JHEP} {\bfseries 11}
  (2021) 110} [\href{https://arxiv.org/abs/2107.03839}{{\ttfamily
  2107.03839}}].

\bibitem{Laporte:2021kyp}
C.~Laporte, A.D.~Pereira, F.~Saueressig and J.~Wang, \emph{{Scalar-tensor
  theories within Asymptotic Safety}},
  \href{https://doi.org/10.1007/JHEP12(2021)001}{\emph{JHEP} {\bfseries 12}
  (2021) 001} [\href{https://arxiv.org/abs/2110.09566}{{\ttfamily
  2110.09566}}].

\bibitem{deBrito:2020dta}
G.P.~de~Brito, A.~Eichhorn and M.~Schiffer, \emph{{Light charged fermions in
  quantum gravity}},
  \href{https://doi.org/10.1016/j.physletb.2021.136128}{\emph{Phys. Lett. B}
  {\bfseries 815} (2021) 136128}
  [\href{https://arxiv.org/abs/2010.00605}{{\ttfamily 2010.00605}}].

\bibitem{Arkani-Hamed:2006emk}
N.~Arkani-Hamed, L.~Motl, A.~Nicolis and C.~Vafa, \emph{{The String landscape,
  black holes and gravity as the weakest force}},
  \href{https://doi.org/10.1088/1126-6708/2007/06/060}{\emph{JHEP} {\bfseries
  06} (2007) 060} [\href{https://arxiv.org/abs/hep-th/0601001}{{\ttfamily
  hep-th/0601001}}].

\bibitem{deAlwis:2019aud}
S.~de~Alwis, A.~Eichhorn, A.~Held, J.M.~Pawlowski, M.~Schiffer and
  F.~Versteegen, \emph{{Asymptotic safety, string theory and the weak gravity
  conjecture}},
  \href{https://doi.org/10.1016/j.physletb.2019.134991}{\emph{Phys. Lett. B}
  {\bfseries 798} (2019) 134991}
  [\href{https://arxiv.org/abs/1907.07894}{{\ttfamily 1907.07894}}].

\bibitem{Basile:2021krr}
I.~Basile and A.~Platania, \emph{{Asymptotic Safety: Swampland or
  Wonderland?}}, \href{https://doi.org/10.3390/universe7100389}{\emph{Universe}
  {\bfseries 7} (2021) 389} [\href{https://arxiv.org/abs/2107.06897}{{\ttfamily
  2107.06897}}].

\bibitem{Vafa:2005ui}
C.~Vafa, \emph{{The String landscape and the swampland}},
  \href{https://arxiv.org/abs/hep-th/0509212}{{\ttfamily hep-th/0509212}}.

\bibitem{Brennan:2017rbf}
T.D.~Brennan, F.~Carta and C.~Vafa, \emph{{The String Landscape, the Swampland,
  and the Missing Corner}},
  \href{https://doi.org/10.22323/1.305.0015}{\emph{PoS} {\bfseries TASI2017}
  (2017) 015} [\href{https://arxiv.org/abs/1711.00864}{{\ttfamily
  1711.00864}}].

\bibitem{Palti:2019pca}
E.~Palti, \emph{{The Swampland: Introduction and Review}},
  \href{https://doi.org/10.1002/prop.201900037}{\emph{Fortsch. Phys.}
  {\bfseries 67} (2019) 1900037}
  [\href{https://arxiv.org/abs/1903.06239}{{\ttfamily 1903.06239}}].

\bibitem{Gockeler:1997dn}
M.~Gockeler, R.~Horsley, V.~Linke, P.E.L.~Rakow, G.~Schierholz and H.~Stuben,
  \emph{{Is there a Landau pole problem in QED?}},
  \href{https://doi.org/10.1103/PhysRevLett.80.4119}{\emph{Phys. Rev. Lett.}
  {\bfseries 80} (1998) 4119}
  [\href{https://arxiv.org/abs/hep-th/9712244}{{\ttfamily hep-th/9712244}}].

\bibitem{Gockeler:1997kt}
M.~Gockeler, R.~Horsley, V.~Linke, P.E.L.~Rakow, G.~Schierholz and H.~Stuben,
  \emph{{Resolution of the Landau pole problem in QED}},
  \href{https://doi.org/10.1016/S0920-5632(97)00875-X}{\emph{Nucl. Phys. B
  Proc. Suppl.} {\bfseries 63} (1998) 694}
  [\href{https://arxiv.org/abs/hep-lat/9801004}{{\ttfamily hep-lat/9801004}}].

\bibitem{Gies:2004hy}
H.~Gies and J.~Jaeckel, \emph{{Renormalization flow of QED}},
  \href{https://doi.org/10.1103/PhysRevLett.93.110405}{\emph{Phys. Rev. Lett.}
  {\bfseries 93} (2004) 110405}
  [\href{https://arxiv.org/abs/hep-ph/0405183}{{\ttfamily hep-ph/0405183}}].

\bibitem{Eichhorn:2017lry}
A.~Eichhorn and F.~Versteegen, \emph{{Upper bound on the Abelian gauge coupling
  from asymptotic safety}},
  \href{https://doi.org/10.1007/JHEP01(2018)030}{\emph{JHEP} {\bfseries 01}
  (2018) 030} [\href{https://arxiv.org/abs/1709.07252}{{\ttfamily
  1709.07252}}].

\bibitem{Daum:2009dn}
J.-E.~Daum, U.~Harst and M.~Reuter, \emph{{Running Gauge Coupling in
  Asymptotically Safe Quantum Gravity}},
  \href{https://doi.org/10.1007/JHEP01(2010)084}{\emph{JHEP} {\bfseries 01}
  (2010) 084} [\href{https://arxiv.org/abs/0910.4938}{{\ttfamily 0910.4938}}].

\bibitem{Harst:2011zx}
U.~Harst and M.~Reuter, \emph{{QED coupled to QEG}},
  \href{https://doi.org/10.1007/JHEP05(2011)119}{\emph{JHEP} {\bfseries 05}
  (2011) 119} [\href{https://arxiv.org/abs/1101.6007}{{\ttfamily 1101.6007}}].

\bibitem{Christiansen:2017cxa}
N.~Christiansen, D.F.~Litim, J.M.~Pawlowski and M.~Reichert, \emph{{Asymptotic
  safety of gravity with matter}},
  \href{https://doi.org/10.1103/PhysRevD.97.106012}{\emph{Phys. Rev. D}
  {\bfseries 97} (2018) 106012}
  [\href{https://arxiv.org/abs/1710.04669}{{\ttfamily 1710.04669}}].

\bibitem{Eichhorn:2018whv}
A.~Eichhorn and A.~Held, \emph{{Mass difference for charged quarks from
  asymptotically safe quantum gravity}},
  \href{https://doi.org/10.1103/PhysRevLett.121.151302}{\emph{Phys. Rev. Lett.}
  {\bfseries 121} (2018) 151302}
  [\href{https://arxiv.org/abs/1803.04027}{{\ttfamily 1803.04027}}].

\bibitem{Folkerts:2011jz}
S.~Folkerts, D.F.~Litim and J.M.~Pawlowski, \emph{{Asymptotic freedom of
  Yang-Mills theory with gravity}},
  \href{https://doi.org/10.1016/j.physletb.2012.02.002}{\emph{Phys. Lett. B}
  {\bfseries 709} (2012) 234}
  [\href{https://arxiv.org/abs/1101.5552}{{\ttfamily 1101.5552}}].

\bibitem{Pietrykowski:2006xy}
A.R.~Pietrykowski, \emph{{Gauge dependence of gravitational correction to
  running of gauge couplings}},
  \href{https://doi.org/10.1103/PhysRevLett.98.061801}{\emph{Phys. Rev. Lett.}
  {\bfseries 98} (2007) 061801}
  [\href{https://arxiv.org/abs/hep-th/0606208}{{\ttfamily hep-th/0606208}}].

\bibitem{Toms:2007sk}
D.J.~Toms, \emph{{Quantum gravity and charge renormalization}},
  \href{https://doi.org/10.1103/PhysRevD.76.045015}{\emph{Phys. Rev. D}
  {\bfseries 76} (2007) 045015}
  [\href{https://arxiv.org/abs/0708.2990}{{\ttfamily 0708.2990}}].

\bibitem{Ebert:2007gf}
D.~Ebert, J.~Plefka and A.~Rodigast, \emph{{Absence of gravitational
  contributions to the running Yang-Mills coupling}},
  \href{https://doi.org/10.1016/j.physletb.2008.01.037}{\emph{Phys. Lett. B}
  {\bfseries 660} (2008) 579}
  [\href{https://arxiv.org/abs/0710.1002}{{\ttfamily 0710.1002}}].

\bibitem{Toms:2010vy}
D.J.~Toms, \emph{{Quantum gravitational contributions to quantum
  electrodynamics}}, \href{https://doi.org/10.1038/nature09506}{\emph{Nature}
  {\bfseries 468} (2010) 56} [\href{https://arxiv.org/abs/1010.0793}{{\ttfamily
  1010.0793}}].

\bibitem{Anber:2010uj}
M.M.~Anber, J.F.~Donoghue and M.~El-Houssieny, \emph{{Running couplings and
  operator mixing in the gravitational corrections to coupling constants}},
  \href{https://doi.org/10.1103/PhysRevD.83.124003}{\emph{Phys. Rev. D}
  {\bfseries 83} (2011) 124003}
  [\href{https://arxiv.org/abs/1011.3229}{{\ttfamily 1011.3229}}].

\bibitem{Eichhorn:2013ug}
A.~Eichhorn, \emph{{Faddeev-Popov ghosts in quantum gravity beyond perturbation
  theory}}, \href{https://doi.org/10.1103/PhysRevD.87.124016}{\emph{Phys. Rev.
  D} {\bfseries 87} (2013) 124016}
  [\href{https://arxiv.org/abs/1301.0632}{{\ttfamily 1301.0632}}].

\bibitem{Eichhorn:2017sok}
A.~Eichhorn, S.~Lippoldt and V.~Skrinjar, \emph{{Nonminimal hints for
  asymptotic safety}},
  \href{https://doi.org/10.1103/PhysRevD.97.026002}{\emph{Phys. Rev. D}
  {\bfseries 97} (2018) 026002}
  [\href{https://arxiv.org/abs/1710.03005}{{\ttfamily 1710.03005}}].

\bibitem{Eichhorn:2011pc}
A.~Eichhorn and H.~Gies, \emph{{Light fermions in quantum gravity}},
  \href{https://doi.org/10.1088/1367-2630/13/12/125012}{\emph{New J. Phys.}
  {\bfseries 13} (2011) 125012}
  [\href{https://arxiv.org/abs/1104.5366}{{\ttfamily 1104.5366}}].

\bibitem{Meibohm:2016mkp}
J.~Meibohm and J.M.~Pawlowski, \emph{{Chiral fermions in asymptotically safe
  quantum gravity}},
  \href{https://doi.org/10.1140/epjc/s10052-016-4132-7}{\emph{Eur. Phys. J. C}
  {\bfseries 76} (2016) 285}
  [\href{https://arxiv.org/abs/1601.04597}{{\ttfamily 1601.04597}}].

\bibitem{Eichhorn:2018nda}
A.~Eichhorn, S.~Lippoldt and M.~Schiffer, \emph{{Zooming in on fermions and
  quantum gravity}},
  \href{https://doi.org/10.1103/PhysRevD.99.086002}{\emph{Phys. Rev. D}
  {\bfseries 99} (2019) 086002}
  [\href{https://arxiv.org/abs/1812.08782}{{\ttfamily 1812.08782}}].

\bibitem{Schiffer:2021gwl}
M.~Schiffer, \emph{{Probing Quantum Gravity: Theoretical and phenomenological
  consistency tests of asymptotically safe quantum gravity}}, Ph.D. thesis, U.
  Heidelberg (main), 2021.
\newblock 10.11588/heidok.00030645.

\bibitem{Wetterich:1992yh}
C.~Wetterich, \emph{{Exact evolution equation for the effective potential}},
  \href{https://doi.org/10.1016/0370-2693(93)90726-X}{\emph{Phys. Lett. B}
  {\bfseries 301} (1993) 90}
  [\href{https://arxiv.org/abs/1710.05815}{{\ttfamily 1710.05815}}].

\bibitem{Ellwanger:1993mw}
U.~Ellwanger, \emph{{FLow equations for N point functions and bound states}},
  \href{https://doi.org/10.1007/BF01555911}{\emph{Z. Phys. C} {\bfseries 62}
  (1994) 503} [\href{https://arxiv.org/abs/hep-ph/9308260}{{\ttfamily
  hep-ph/9308260}}].

\bibitem{Morris:1993qb}
T.R.~Morris, \emph{{The Exact renormalization group and approximate
  solutions}}, \href{https://doi.org/10.1142/S0217751X94000972}{\emph{Int. J.
  Mod. Phys. A} {\bfseries 9} (1994) 2411}
  [\href{https://arxiv.org/abs/hep-ph/9308265}{{\ttfamily hep-ph/9308265}}].

\bibitem{Reuter:1996cp}
M.~Reuter, \emph{{Nonperturbative evolution equation for quantum gravity}},
  \href{https://doi.org/10.1103/PhysRevD.57.971}{\emph{Phys. Rev. D} {\bfseries
  57} (1998) 971} [\href{https://arxiv.org/abs/hep-th/9605030}{{\ttfamily
  hep-th/9605030}}].

\bibitem{Berges:2000ew}
J.~Berges, N.~Tetradis and C.~Wetterich, \emph{{Nonperturbative renormalization
  flow in quantum field theory and statistical physics}},
  \href{https://doi.org/10.1016/S0370-1573(01)00098-9}{\emph{Phys. Rept.}
  {\bfseries 363} (2002) 223}
  [\href{https://arxiv.org/abs/hep-ph/0005122}{{\ttfamily hep-ph/0005122}}].

\bibitem{Pawlowski:2005xe}
J.M.~Pawlowski, \emph{{Aspects of the functional renormalisation group}},
  \href{https://doi.org/10.1016/j.aop.2007.01.007}{\emph{Annals Phys.}
  {\bfseries 322} (2007) 2831}
  [\href{https://arxiv.org/abs/hep-th/0512261}{{\ttfamily hep-th/0512261}}].

\bibitem{Gies:2006wv}
H.~Gies, \emph{{Introduction to the functional RG and applications to gauge
  theories}}, \href{https://doi.org/10.1007/978-3-642-27320-9_6}{\emph{Lect.
  Notes Phys.} {\bfseries 852} (2012) 287}
  [\href{https://arxiv.org/abs/hep-ph/0611146}{{\ttfamily hep-ph/0611146}}].

\bibitem{Delamotte:2007pf}
B.~Delamotte, \emph{{An Introduction to the nonperturbative renormalization
  group}}, \href{https://doi.org/10.1007/978-3-642-27320-9_2}{\emph{Lect. Notes
  Phys.} {\bfseries 852} (2012) 49}
  [\href{https://arxiv.org/abs/cond-mat/0702365}{{\ttfamily
  cond-mat/0702365}}].

\bibitem{Rosten:2010vm}
O.J.~Rosten, \emph{{Fundamentals of the Exact Renormalization Group}},
  \href{https://doi.org/10.1016/j.physrep.2011.12.003}{\emph{Phys. Rept.}
  {\bfseries 511} (2012) 177}
  [\href{https://arxiv.org/abs/1003.1366}{{\ttfamily 1003.1366}}].

\bibitem{Braun:2011pp}
J.~Braun, \emph{{Fermion Interactions and Universal Behavior in Strongly
  Interacting Theories}},
  \href{https://doi.org/10.1088/0954-3899/39/3/033001}{\emph{J. Phys. G}
  {\bfseries 39} (2012) 033001}
  [\href{https://arxiv.org/abs/1108.4449}{{\ttfamily 1108.4449}}].

\bibitem{Reuter:2012id}
M.~Reuter and F.~Saueressig, \emph{{Quantum Einstein Gravity}},
  \href{https://doi.org/10.1088/1367-2630/14/5/055022}{\emph{New J. Phys.}
  {\bfseries 14} (2012) 055022}
  [\href{https://arxiv.org/abs/1202.2274}{{\ttfamily 1202.2274}}].

\bibitem{Dupuis:2020fhh}
N.~Dupuis, L.~Canet, A.~Eichhorn, W.~Metzner, J.M.~Pawlowski, M.~Tissier
  et~al., \emph{{The nonperturbative functional renormalization group and its
  applications}},
  \href{https://doi.org/10.1016/j.physrep.2021.01.001}{\emph{Phys. Rept.}
  {\bfseries 910} (2021) 1} [\href{https://arxiv.org/abs/2006.04853}{{\ttfamily
  2006.04853}}].

\bibitem{Gies:2002af}
H.~Gies, \emph{{Running coupling in Yang-Mills theory: A flow equation study}},
  \href{https://doi.org/10.1103/PhysRevD.66.025006}{\emph{Phys. Rev. D}
  {\bfseries 66} (2002) 025006}
  [\href{https://arxiv.org/abs/hep-th/0202207}{{\ttfamily hep-th/0202207}}].

\bibitem{Benedetti:2010nr}
D.~Benedetti, K.~Groh, P.F.~Machado and F.~Saueressig, \emph{{The Universal RG
  Machine}}, \href{https://doi.org/10.1007/JHEP06(2011)079}{\emph{JHEP}
  {\bfseries 06} (2011) 079} [\href{https://arxiv.org/abs/1012.3081}{{\ttfamily
  1012.3081}}].

\bibitem{Gies:2015tca}
H.~Gies, B.~Knorr and S.~Lippoldt, \emph{{Generalized Parametrization
  Dependence in Quantum Gravity}},
  \href{https://doi.org/10.1103/PhysRevD.92.084020}{\emph{Phys. Rev. D}
  {\bfseries 92} (2015) 084020}
  [\href{https://arxiv.org/abs/1507.08859}{{\ttfamily 1507.08859}}].

\bibitem{Litim:2001up}
D.F.~Litim, \emph{{Optimized renormalization group flows}},
  \href{https://doi.org/10.1103/PhysRevD.64.105007}{\emph{Phys. Rev. D}
  {\bfseries 64} (2001) 105007}
  [\href{https://arxiv.org/abs/hep-th/0103195}{{\ttfamily hep-th/0103195}}].

\bibitem{Knorr:2021slg}
B.~Knorr, \emph{{The derivative expansion in asymptotically safe quantum
  gravity: general setup and quartic order}},
  \href{https://doi.org/10.21468/SciPostPhysCore.4.3.020}{\emph{SciPost Phys.}
  {\bfseries 4} (2021) 020} [\href{https://arxiv.org/abs/2104.11336}{{\ttfamily
  2104.11336}}].

\bibitem{Heisenberg:1936nmg}
W.~Heisenberg and H.~Euler, \emph{{Consequences of Dirac's theory of
  positrons}}, \href{https://doi.org/10.1007/BF01343663}{\emph{Z. Phys.}
  {\bfseries 98} (1936) 714}
  [\href{https://arxiv.org/abs/physics/0605038}{{\ttfamily physics/0605038}}].

\bibitem{Denz:2016qks}
T.~Denz, J.M.~Pawlowski and M.~Reichert, \emph{{Towards apparent convergence in
  asymptotically safe quantum gravity}},
  \href{https://doi.org/10.1140/epjc/s10052-018-5806-0}{\emph{Eur. Phys. J. C}
  {\bfseries 78} (2018) 336}
  [\href{https://arxiv.org/abs/1612.07315}{{\ttfamily 1612.07315}}].

\bibitem{Eichhorn:2018akn}
A.~Eichhorn, P.~Labus, J.M.~Pawlowski and M.~Reichert, \emph{{Effective
  universality in quantum gravity}},
  \href{https://doi.org/10.21468/SciPostPhys.5.4.031}{\emph{SciPost Phys.}
  {\bfseries 5} (2018) 031} [\href{https://arxiv.org/abs/1804.00012}{{\ttfamily
  1804.00012}}].

\bibitem{Eichhorn:2018ydy}
A.~Eichhorn, S.~Lippoldt, J.M.~Pawlowski, M.~Reichert and M.~Schiffer,
  \emph{{How perturbative is quantum gravity?}},
  \href{https://doi.org/10.1016/j.physletb.2019.01.071}{\emph{Phys. Lett. B}
  {\bfseries 792} (2019) 310}
  [\href{https://arxiv.org/abs/1810.02828}{{\ttfamily 1810.02828}}].

\bibitem{Knorr:2021niv}
B.~Knorr and M.~Schiffer, \emph{{Non-Perturbative Propagators in Quantum
  Gravity}}, \href{https://doi.org/10.3390/universe7070216}{\emph{Universe}
  {\bfseries 7} (2021) 216} [\href{https://arxiv.org/abs/2105.04566}{{\ttfamily
  2105.04566}}].

\bibitem{Brizuela:2008ra}
D.~Brizuela, J.M.~Martin-Garcia and G.A.~Mena~Marugan, \emph{{xPert: Computer
  algebra for metric perturbation theory}},
  \href{https://doi.org/10.1007/s10714-009-0773-2}{\emph{Gen. Rel. Grav.}
  {\bfseries 41} (2009) 2415}
  [\href{https://arxiv.org/abs/0807.0824}{{\ttfamily 0807.0824}}].

\bibitem{Martin-Garcia:2007bqa}
J.M.~Martin-Garcia, R.~Portugal and L.R.U.~Manssur, \emph{{The Invar Tensor
  Package}}, \href{https://doi.org/10.1016/j.cpc.2007.05.015}{\emph{Comput.
  Phys. Commun.} {\bfseries 177} (2007) 640}
  [\href{https://arxiv.org/abs/0704.1756}{{\ttfamily 0704.1756}}].

\bibitem{Martin-Garcia:2008yei}
J.M.~Martin-Garcia, D.~Yllanes and R.~Portugal, \emph{{The Invar tensor
  package: Differential invariants of Riemann}},
  \href{https://doi.org/10.1016/j.cpc.2008.04.018}{\emph{Comput. Phys. Commun.}
  {\bfseries 179} (2008) 586}
  [\href{https://arxiv.org/abs/0802.1274}{{\ttfamily 0802.1274}}].

\bibitem{2008CoPhC.179..597M}
J.M.~{Mart{\'{\i}}n-Garc{\'{\i}}a}, \emph{{xPerm: fast index canonicalization
  for tensor computer algebra}},
  \href{https://doi.org/10.1016/j.cpc.2008.05.009}{\emph{Computer Physics
  Communications} {\bfseries 179} (2008) 597}
  [\href{https://arxiv.org/abs/0803.0862}{{\ttfamily 0803.0862}}].

\bibitem{2014CoPhC.185.1719N}
T.~{Nutma}, \emph{{xTras: A field-theory inspired xAct package for
  mathematica}},
  \href{https://doi.org/10.1016/j.cpc.2014.02.006}{\emph{Computer Physics
  Communications} {\bfseries 185} (2014) 1719}
  [\href{https://arxiv.org/abs/1308.3493}{{\ttfamily 1308.3493}}].

\bibitem{Cyrol:2016zqb}
A.K.~Cyrol, M.~Mitter and N.~Strodthoff, \emph{{FormTracer - A Mathematica
  Tracing Package Using FORM}},
  \href{https://doi.org/10.1016/j.cpc.2017.05.024}{\emph{Comput. Phys. Commun.}
  {\bfseries 219} (2017) 346}
  [\href{https://arxiv.org/abs/1610.09331}{{\ttfamily 1610.09331}}].

\bibitem{Dittrich:2000zu}
W.~Dittrich and H.~Gies, \emph{{Probing the quantum vacuum. Perturbative
  effective action approach in quantum electrodynamics and its application}},
  vol.~166 (2000),
  \href{https://doi.org/10.1007/3-540-45585-X}{10.1007/3-540-45585-X}.

\bibitem{Knorr:2017kye}
B.~Knorr, \emph{{Asymptotic safety in QFT: from quantum gravity to graphene}},
  Ph.D. thesis, Jena U., 2017.

\bibitem{Fabbrichesi:2020wbt}
M.~Fabbrichesi, E.~Gabrielli and G.~Lanfranchi, \emph{{The Dark Photon}},
  \href{https://arxiv.org/abs/2005.01515}{{\ttfamily 2005.01515}}.

\bibitem{Eichhorn:2020mte}
A.~Eichhorn, \emph{{Asymptotically safe gravity}},  in \emph{{57th
  International School of Subnuclear Physics}: {In Search for the Unexpected}},
  2, 2020 [\href{https://arxiv.org/abs/2003.00044}{{\ttfamily 2003.00044}}].

\bibitem{Percacci:2010af}
R.~Percacci and G.P.~Vacca, \emph{{Asymptotic Safety, Emergence and Minimal
  Length}}, \href{https://doi.org/10.1088/0264-9381/27/24/245026}{\emph{Class.
  Quant. Grav.} {\bfseries 27} (2010) 245026}
  [\href{https://arxiv.org/abs/1008.3621}{{\ttfamily 1008.3621}}].

\bibitem{Held:2020kze}
A.~Held, \emph{{Effective asymptotic safety and its predictive power:
  Gauge-Yukawa theories}},
  \href{https://doi.org/10.3389/fphy.2020.00341}{\emph{Front. in Phys.}
  {\bfseries 8} (2020) 341} [\href{https://arxiv.org/abs/2003.13642}{{\ttfamily
  2003.13642}}].

\bibitem{Souma:1999at}
W.~Souma, \emph{{Nontrivial ultraviolet fixed point in quantum gravity}},
  \href{https://doi.org/10.1143/PTP.102.181}{\emph{Prog. Theor. Phys.}
  {\bfseries 102} (1999) 181}
  [\href{https://arxiv.org/abs/hep-th/9907027}{{\ttfamily hep-th/9907027}}].

\bibitem{Lauscher:2001ya}
O.~Lauscher and M.~Reuter, \emph{{Ultraviolet fixed point and generalized flow
  equation of quantum gravity}},
  \href{https://doi.org/10.1103/PhysRevD.65.025013}{\emph{Phys. Rev. D}
  {\bfseries 65} (2002) 025013}
  [\href{https://arxiv.org/abs/hep-th/0108040}{{\ttfamily hep-th/0108040}}].

\bibitem{Reuter:2001ag}
M.~Reuter and F.~Saueressig, \emph{{Renormalization group flow of quantum
  gravity in the Einstein-Hilbert truncation}},
  \href{https://doi.org/10.1103/PhysRevD.65.065016}{\emph{Phys. Rev. D}
  {\bfseries 65} (2002) 065016}
  [\href{https://arxiv.org/abs/hep-th/0110054}{{\ttfamily hep-th/0110054}}].

\bibitem{Lauscher:2002sq}
O.~Lauscher and M.~Reuter, \emph{{Flow equation of quantum Einstein gravity in
  a higher derivative truncation}},
  \href{https://doi.org/10.1103/PhysRevD.66.025026}{\emph{Phys. Rev. D}
  {\bfseries 66} (2002) 025026}
  [\href{https://arxiv.org/abs/hep-th/0205062}{{\ttfamily hep-th/0205062}}].

\bibitem{Litim:2003vp}
D.F.~Litim, \emph{{Fixed points of quantum gravity}},
  \href{https://doi.org/10.1103/PhysRevLett.92.201301}{\emph{Phys. Rev. Lett.}
  {\bfseries 92} (2004) 201301}
  [\href{https://arxiv.org/abs/hep-th/0312114}{{\ttfamily hep-th/0312114}}].

\bibitem{Codello:2006in}
A.~Codello and R.~Percacci, \emph{{Fixed points of higher derivative gravity}},
  \href{https://doi.org/10.1103/PhysRevLett.97.221301}{\emph{Phys. Rev. Lett.}
  {\bfseries 97} (2006) 221301}
  [\href{https://arxiv.org/abs/hep-th/0607128}{{\ttfamily hep-th/0607128}}].

\bibitem{Machado:2007ea}
P.F.~Machado and F.~Saueressig, \emph{{On the renormalization group flow of
  f(R)-gravity}}, \href{https://doi.org/10.1103/PhysRevD.77.124045}{\emph{Phys.
  Rev. D} {\bfseries 77} (2008) 124045}
  [\href{https://arxiv.org/abs/0712.0445}{{\ttfamily 0712.0445}}].

\bibitem{Codello:2008vh}
A.~Codello, R.~Percacci and C.~Rahmede, \emph{{Investigating the Ultraviolet
  Properties of Gravity with a Wilsonian Renormalization Group Equation}},
  \href{https://doi.org/10.1016/j.aop.2008.08.008}{\emph{Annals Phys.}
  {\bfseries 324} (2009) 414}
  [\href{https://arxiv.org/abs/0805.2909}{{\ttfamily 0805.2909}}].

\bibitem{Benedetti:2009rx}
D.~Benedetti, P.F.~Machado and F.~Saueressig, \emph{{Asymptotic safety in
  higher-derivative gravity}},
  \href{https://doi.org/10.1142/S0217732309031521}{\emph{Mod. Phys. Lett. A}
  {\bfseries 24} (2009) 2233}
  [\href{https://arxiv.org/abs/0901.2984}{{\ttfamily 0901.2984}}].

\bibitem{Eichhorn:2009ah}
A.~Eichhorn, H.~Gies and M.M.~Scherer, \emph{{Asymptotically free scalar
  curvature-ghost coupling in Quantum Einstein Gravity}},
  \href{https://doi.org/10.1103/PhysRevD.80.104003}{\emph{Phys. Rev. D}
  {\bfseries 80} (2009) 104003}
  [\href{https://arxiv.org/abs/0907.1828}{{\ttfamily 0907.1828}}].

\bibitem{Manrique:2010am}
E.~Manrique, M.~Reuter and F.~Saueressig, \emph{{Bimetric Renormalization Group
  Flows in Quantum Einstein Gravity}},
  \href{https://doi.org/10.1016/j.aop.2010.11.006}{\emph{Annals Phys.}
  {\bfseries 326} (2011) 463}
  [\href{https://arxiv.org/abs/1006.0099}{{\ttfamily 1006.0099}}].

\bibitem{Eichhorn:2010tb}
A.~Eichhorn and H.~Gies, \emph{{Ghost anomalous dimension in asymptotically
  safe quantum gravity}},
  \href{https://doi.org/10.1103/PhysRevD.81.104010}{\emph{Phys. Rev. D}
  {\bfseries 81} (2010) 104010}
  [\href{https://arxiv.org/abs/1001.5033}{{\ttfamily 1001.5033}}].

\bibitem{Groh:2010ta}
K.~Groh and F.~Saueressig, \emph{{Ghost wave-function renormalization in
  Asymptotically Safe Quantum Gravity}},
  \href{https://doi.org/10.1088/1751-8113/43/36/365403}{\emph{J. Phys. A}
  {\bfseries 43} (2010) 365403}
  [\href{https://arxiv.org/abs/1001.5032}{{\ttfamily 1001.5032}}].

\bibitem{Dietz:2012ic}
J.A.~Dietz and T.R.~Morris, \emph{{Asymptotic safety in the f(R)
  approximation}}, \href{https://doi.org/10.1007/JHEP01(2013)108}{\emph{JHEP}
  {\bfseries 01} (2013) 108} [\href{https://arxiv.org/abs/1211.0955}{{\ttfamily
  1211.0955}}].

\bibitem{Christiansen:2012rx}
N.~Christiansen, D.F.~Litim, J.M.~Pawlowski and A.~Rodigast, \emph{{Fixed
  points and infrared completion of quantum gravity}},
  \href{https://doi.org/10.1016/j.physletb.2013.11.025}{\emph{Phys. Lett. B}
  {\bfseries 728} (2014) 114}
  [\href{https://arxiv.org/abs/1209.4038}{{\ttfamily 1209.4038}}].

\bibitem{Rechenberger:2012pm}
S.~Rechenberger and F.~Saueressig, \emph{{The $R^2$ phase-diagram of QEG and
  its spectral dimension}},
  \href{https://doi.org/10.1103/PhysRevD.86.024018}{\emph{Phys. Rev. D}
  {\bfseries 86} (2012) 024018}
  [\href{https://arxiv.org/abs/1206.0657}{{\ttfamily 1206.0657}}].

\bibitem{Falls:2013bv}
K.~Falls, D.F.~Litim, K.~Nikolakopoulos and C.~Rahmede, \emph{{A bootstrap
  towards asymptotic safety}},
  \href{https://arxiv.org/abs/1301.4191}{{\ttfamily 1301.4191}}.

\bibitem{Ohta:2013uca}
N.~Ohta and R.~Percacci, \emph{{Higher Derivative Gravity and Asymptotic Safety
  in Diverse Dimensions}},
  \href{https://doi.org/10.1088/0264-9381/31/1/015024}{\emph{Class. Quant.
  Grav.} {\bfseries 31} (2014) 015024}
  [\href{https://arxiv.org/abs/1308.3398}{{\ttfamily 1308.3398}}].

\bibitem{Eichhorn:2013xr}
A.~Eichhorn, \emph{{On unimodular quantum gravity}},
  \href{https://doi.org/10.1088/0264-9381/30/11/115016}{\emph{Class. Quant.
  Grav.} {\bfseries 30} (2013) 115016}
  [\href{https://arxiv.org/abs/1301.0879}{{\ttfamily 1301.0879}}].

\bibitem{Falls:2014tra}
K.~Falls, D.F.~Litim, K.~Nikolakopoulos and C.~Rahmede, \emph{{Further evidence
  for asymptotic safety of quantum gravity}},
  \href{https://doi.org/10.1103/PhysRevD.93.104022}{\emph{Phys. Rev. D}
  {\bfseries 93} (2016) 104022}
  [\href{https://arxiv.org/abs/1410.4815}{{\ttfamily 1410.4815}}].

\bibitem{Codello:2013fpa}
A.~Codello, G.~D'Odorico and C.~Pagani, \emph{{Consistent closure of
  renormalization group flow equations in quantum gravity}},
  \href{https://doi.org/10.1103/PhysRevD.89.081701}{\emph{Phys. Rev. D}
  {\bfseries 89} (2014) 081701}
  [\href{https://arxiv.org/abs/1304.4777}{{\ttfamily 1304.4777}}].

\bibitem{Christiansen:2014raa}
N.~Christiansen, B.~Knorr, J.M.~Pawlowski and A.~Rodigast, \emph{{Global Flows
  in Quantum Gravity}},
  \href{https://doi.org/10.1103/PhysRevD.93.044036}{\emph{Phys. Rev. D}
  {\bfseries 93} (2016) 044036}
  [\href{https://arxiv.org/abs/1403.1232}{{\ttfamily 1403.1232}}].

\bibitem{Demmel:2015oqa}
M.~Demmel, F.~Saueressig and O.~Zanusso, \emph{{A proper fixed functional for
  four-dimensional Quantum Einstein Gravity}},
  \href{https://doi.org/10.1007/JHEP08(2015)113}{\emph{JHEP} {\bfseries 08}
  (2015) 113} [\href{https://arxiv.org/abs/1504.07656}{{\ttfamily
  1504.07656}}].

\bibitem{Christiansen:2015rva}
N.~Christiansen, B.~Knorr, J.~Meibohm, J.M.~Pawlowski and M.~Reichert,
  \emph{{Local Quantum Gravity}},
  \href{https://doi.org/10.1103/PhysRevD.92.121501}{\emph{Phys. Rev. D}
  {\bfseries 92} (2015) 121501}
  [\href{https://arxiv.org/abs/1506.07016}{{\ttfamily 1506.07016}}].

\bibitem{Ohta:2015fcu}
N.~Ohta, R.~Percacci and G.P.~Vacca, \emph{{Renormalization Group Equation and
  scaling solutions for f(R) gravity in exponential parametrization}},
  \href{https://doi.org/10.1140/epjc/s10052-016-3895-1}{\emph{Eur. Phys. J. C}
  {\bfseries 76} (2016) 46} [\href{https://arxiv.org/abs/1511.09393}{{\ttfamily
  1511.09393}}].

\bibitem{Ohta:2015efa}
N.~Ohta, R.~Percacci and G.P.~Vacca, \emph{{Flow equation for $f(R)$ gravity
  and some of its exact solutions}},
  \href{https://doi.org/10.1103/PhysRevD.92.061501}{\emph{Phys. Rev. D}
  {\bfseries 92} (2015) 061501}
  [\href{https://arxiv.org/abs/1507.00968}{{\ttfamily 1507.00968}}].

\bibitem{Falls:2015qga}
K.~Falls, \emph{{Renormalization of Newton\textquoteright{}s constant}},
  \href{https://doi.org/10.1103/PhysRevD.92.124057}{\emph{Phys. Rev. D}
  {\bfseries 92} (2015) 124057}
  [\href{https://arxiv.org/abs/1501.05331}{{\ttfamily 1501.05331}}].

\bibitem{Eichhorn:2015bna}
A.~Eichhorn, \emph{{The Renormalization Group flow of unimodular f(R)
  gravity}}, \href{https://doi.org/10.1007/JHEP04(2015)096}{\emph{JHEP}
  {\bfseries 04} (2015) 096}
  [\href{https://arxiv.org/abs/1501.05848}{{\ttfamily 1501.05848}}].

\bibitem{Gies:2016con}
H.~Gies, B.~Knorr, S.~Lippoldt and F.~Saueressig, \emph{{Gravitational Two-Loop
  Counterterm Is Asymptotically Safe}},
  \href{https://doi.org/10.1103/PhysRevLett.116.211302}{\emph{Phys. Rev. Lett.}
  {\bfseries 116} (2016) 211302}
  [\href{https://arxiv.org/abs/1601.01800}{{\ttfamily 1601.01800}}].

\bibitem{Biemans:2016rvp}
J.~Biemans, A.~Platania and F.~Saueressig, \emph{{Quantum gravity on foliated
  spacetimes: Asymptotically safe and sound}},
  \href{https://doi.org/10.1103/PhysRevD.95.086013}{\emph{Phys. Rev. D}
  {\bfseries 95} (2017) 086013}
  [\href{https://arxiv.org/abs/1609.04813}{{\ttfamily 1609.04813}}].

\bibitem{Falls:2016msz}
K.~Falls and N.~Ohta, \emph{{Renormalization Group Equation for $f(R)$ gravity
  on hyperbolic spaces}},
  \href{https://doi.org/10.1103/PhysRevD.94.084005}{\emph{Phys. Rev. D}
  {\bfseries 94} (2016) 084005}
  [\href{https://arxiv.org/abs/1607.08460}{{\ttfamily 1607.08460}}].

\bibitem{Falls:2016wsa}
K.~Falls, D.F.~Litim, K.~Nikolakopoulos and C.~Rahmede, \emph{{On de Sitter
  solutions in asymptotically safe $f(R)$ theories}},
  \href{https://doi.org/10.1088/1361-6382/aac440}{\emph{Class. Quant. Grav.}
  {\bfseries 35} (2018) 135006}
  [\href{https://arxiv.org/abs/1607.04962}{{\ttfamily 1607.04962}}].

\bibitem{deAlwis:2017ysy}
S.P.~de~Alwis, \emph{{Exact RG Flow Equations and Quantum Gravity}},
  \href{https://doi.org/10.1007/JHEP03(2018)118}{\emph{JHEP} {\bfseries 03}
  (2018) 118} [\href{https://arxiv.org/abs/1707.09298}{{\ttfamily
  1707.09298}}].

\bibitem{Christiansen:2017bsy}
N.~Christiansen, K.~Falls, J.M.~Pawlowski and M.~Reichert, \emph{{Curvature
  dependence of quantum gravity}},
  \href{https://doi.org/10.1103/PhysRevD.97.046007}{\emph{Phys. Rev. D}
  {\bfseries 97} (2018) 046007}
  [\href{https://arxiv.org/abs/1711.09259}{{\ttfamily 1711.09259}}].

\bibitem{Falls:2017lst}
K.~Falls, C.R.~King, D.F.~Litim, K.~Nikolakopoulos and C.~Rahmede,
  \emph{{Asymptotic safety of quantum gravity beyond Ricci scalars}},
  \href{https://doi.org/10.1103/PhysRevD.97.086006}{\emph{Phys. Rev. D}
  {\bfseries 97} (2018) 086006}
  [\href{https://arxiv.org/abs/1801.00162}{{\ttfamily 1801.00162}}].

\bibitem{Houthoff:2017oam}
W.B.~Houthoff, A.~Kurov and F.~Saueressig, \emph{{Impact of topology in
  foliated Quantum Einstein Gravity}},
  \href{https://doi.org/10.1140/epjc/s10052-017-5046-8}{\emph{Eur. Phys. J. C}
  {\bfseries 77} (2017) 491}
  [\href{https://arxiv.org/abs/1705.01848}{{\ttfamily 1705.01848}}].

\bibitem{Falls:2017cze}
K.~Falls, \emph{{Physical renormalization schemes and asymptotic safety in
  quantum gravity}},
  \href{https://doi.org/10.1103/PhysRevD.96.126016}{\emph{Phys. Rev. D}
  {\bfseries 96} (2017) 126016}
  [\href{https://arxiv.org/abs/1702.03577}{{\ttfamily 1702.03577}}].

\bibitem{Becker:2017tcx}
D.~Becker, C.~Ripken and F.~Saueressig, \emph{{On avoiding Ostrogradski
  instabilities within Asymptotic Safety}},
  \href{https://doi.org/10.1007/JHEP12(2017)121}{\emph{JHEP} {\bfseries 12}
  (2017) 121} [\href{https://arxiv.org/abs/1709.09098}{{\ttfamily
  1709.09098}}].

\bibitem{Knorr:2017fus}
B.~Knorr and S.~Lippoldt, \emph{{Correlation functions on a curved
  background}}, \href{https://doi.org/10.1103/PhysRevD.96.065020}{\emph{Phys.
  Rev. D} {\bfseries 96} (2017) 065020}
  [\href{https://arxiv.org/abs/1707.01397}{{\ttfamily 1707.01397}}].

\bibitem{Knorr:2017mhu}
B.~Knorr, \emph{{Infinite order quantum-gravitational correlations}},
  \href{https://doi.org/10.1088/1361-6382/aabaa0}{\emph{Class. Quant. Grav.}
  {\bfseries 35} (2018) 115005}
  [\href{https://arxiv.org/abs/1710.07055}{{\ttfamily 1710.07055}}].

\bibitem{DeBrito:2018hur}
G.P.~De~Brito, N.~Ohta, A.D.~Pereira, A.A.~Tomaz and M.~Yamada,
  \emph{{Asymptotic safety and field parametrization dependence in the $f(R)$
  truncation}}, \href{https://doi.org/10.1103/PhysRevD.98.026027}{\emph{Phys.
  Rev. D} {\bfseries 98} (2018) 026027}
  [\href{https://arxiv.org/abs/1805.09656}{{\ttfamily 1805.09656}}].

\bibitem{Falls:2018ylp}
K.G.~Falls, D.F.~Litim and J.~Schr\"oder, \emph{{Aspects of asymptotic safety
  for quantum gravity}},
  \href{https://doi.org/10.1103/PhysRevD.99.126015}{\emph{Phys. Rev. D}
  {\bfseries 99} (2019) 126015}
  [\href{https://arxiv.org/abs/1810.08550}{{\ttfamily 1810.08550}}].

\bibitem{Bosma:2019aiu}
L.~Bosma, B.~Knorr and F.~Saueressig, \emph{{Resolving Spacetime Singularities
  within Asymptotic Safety}},
  \href{https://doi.org/10.1103/PhysRevLett.123.101301}{\emph{Phys. Rev. Lett.}
  {\bfseries 123} (2019) 101301}
  [\href{https://arxiv.org/abs/1904.04845}{{\ttfamily 1904.04845}}].

\bibitem{Knorr:2019atm}
B.~Knorr, C.~Ripken and F.~Saueressig, \emph{{Form Factors in Asymptotic
  Safety: conceptual ideas and computational toolbox}},
  \href{https://doi.org/10.1088/1361-6382/ab4a53}{\emph{Class. Quant. Grav.}
  {\bfseries 36} (2019) 234001}
  [\href{https://arxiv.org/abs/1907.02903}{{\ttfamily 1907.02903}}].

\bibitem{Falls:2020qhj}
K.~Falls, N.~Ohta and R.~Percacci, \emph{{Towards the determination of the
  dimension of the critical surface in asymptotically safe gravity}},
  \href{https://doi.org/10.1016/j.physletb.2020.135773}{\emph{Phys. Lett. B}
  {\bfseries 810} (2020) 135773}
  [\href{https://arxiv.org/abs/2004.04126}{{\ttfamily 2004.04126}}].

\bibitem{Kluth:2020bdv}
Y.~Kluth and D.F.~Litim, \emph{{Fixed Points of Quantum Gravity and the
  Dimensionality of the UV Critical Surface}},
  \href{https://arxiv.org/abs/2008.09181}{{\ttfamily 2008.09181}}.

\bibitem{Bonanno:2021squ}
A.~Bonanno, T.~Denz, J.M.~Pawlowski and M.~Reichert, \emph{{Reconstructing the
  graviton}},  \href{https://arxiv.org/abs/2102.02217}{{\ttfamily 2102.02217}}.

\bibitem{Baldazzi:2021orb}
A.~Baldazzi and K.~Falls, \emph{{Essential Quantum Einstein Gravity}},
  \href{https://doi.org/10.3390/universe7080294}{\emph{Universe} {\bfseries 7}
  (2021) 294} [\href{https://arxiv.org/abs/2107.00671}{{\ttfamily
  2107.00671}}].

\bibitem{Sen:2021ffc}
S.~Sen, C.~Wetterich and M.~Yamada, \emph{{Asymptotic freedom and safety in
  quantum gravity}},  \href{https://arxiv.org/abs/2111.04696}{{\ttfamily
  2111.04696}}.

\bibitem{Mitchell:2021qjr}
A.~Mitchell, T.R.~Morris and D.~Stulga, \emph{{Provable properties of
  asymptotic safety in $f(R)$ approximation}},
  \href{https://arxiv.org/abs/2111.05067}{{\ttfamily 2111.05067}}.

\bibitem{Knorr:2021iwv}
B.~Knorr, C.~Ripken and F.~Saueressig, \emph{{Form Factors in Quantum Gravity -
  contrasting non-local, ghost-free gravity and Asymptotic Safety}},
  \href{https://arxiv.org/abs/2111.12365}{{\ttfamily 2111.12365}}.

\bibitem{Baldazzi:2021fye}
A.~Baldazzi, K.~Falls and R.~Ferrero, \emph{{Relational observables in
  Asymptotically safe gravity}},
  \href{https://arxiv.org/abs/2112.02118}{{\ttfamily 2112.02118}}.

\bibitem{Fehre:2021eob}
J.~Fehre, D.F.~Litim, J.M.~Pawlowski and M.~Reichert, \emph{{Lorentzian quantum
  gravity and the graviton spectral function}},
  \href{https://arxiv.org/abs/2111.13232}{{\ttfamily 2111.13232}}.

\bibitem{Narain:2009fy}
G.~Narain and R.~Percacci, \emph{{Renormalization Group Flow in Scalar-Tensor
  Theories. I}},
  \href{https://doi.org/10.1088/0264-9381/27/7/075001}{\emph{Class. Quant.
  Grav.} {\bfseries 27} (2010) 075001}
  [\href{https://arxiv.org/abs/0911.0386}{{\ttfamily 0911.0386}}].

\bibitem{Dona:2012am}
P.~Dona and R.~Percacci, \emph{{Functional renormalization with fermions and
  tetrads}}, \href{https://doi.org/10.1103/PhysRevD.87.045002}{\emph{Phys. Rev.
  D} {\bfseries 87} (2013) 045002}
  [\href{https://arxiv.org/abs/1209.3649}{{\ttfamily 1209.3649}}].

\bibitem{Percacci:2015wwa}
R.~Percacci and G.P.~Vacca, \emph{{Search of scaling solutions in scalar-tensor
  gravity}}, \href{https://doi.org/10.1140/epjc/s10052-015-3410-0}{\emph{Eur.
  Phys. J. C} {\bfseries 75} (2015) 188}
  [\href{https://arxiv.org/abs/1501.00888}{{\ttfamily 1501.00888}}].

\bibitem{Labus:2015ska}
P.~Labus, R.~Percacci and G.P.~Vacca, \emph{{Asymptotic safety in $O(N)$ scalar
  models coupled to gravity}},
  \href{https://doi.org/10.1016/j.physletb.2015.12.022}{\emph{Phys. Lett. B}
  {\bfseries 753} (2016) 274}
  [\href{https://arxiv.org/abs/1505.05393}{{\ttfamily 1505.05393}}].

\bibitem{Dona:2013qba}
P.~Don\`a, A.~Eichhorn and R.~Percacci, \emph{{Matter matters in asymptotically
  safe quantum gravity}},
  \href{https://doi.org/10.1103/PhysRevD.89.084035}{\emph{Phys. Rev. D}
  {\bfseries 89} (2014) 084035}
  [\href{https://arxiv.org/abs/1311.2898}{{\ttfamily 1311.2898}}].

\bibitem{Meibohm:2015twa}
J.~Meibohm, J.M.~Pawlowski and M.~Reichert, \emph{{Asymptotic safety of
  gravity-matter systems}},
  \href{https://doi.org/10.1103/PhysRevD.93.084035}{\emph{Phys. Rev. D}
  {\bfseries 93} (2016) 084035}
  [\href{https://arxiv.org/abs/1510.07018}{{\ttfamily 1510.07018}}].

\bibitem{Dona:2015tnf}
P.~Don\`a, A.~Eichhorn, P.~Labus and R.~Percacci, \emph{{Asymptotic safety in
  an interacting system of gravity and scalar matter}},
  \href{https://doi.org/10.1103/PhysRevD.93.129904}{\emph{Phys. Rev. D}
  {\bfseries 93} (2016) 044049}
  [\href{https://arxiv.org/abs/1512.01589}{{\ttfamily 1512.01589}}].

\bibitem{Biemans:2017zca}
J.~Biemans, A.~Platania and F.~Saueressig, \emph{{Renormalization group fixed
  points of foliated gravity-matter systems}},
  \href{https://doi.org/10.1007/JHEP05(2017)093}{\emph{JHEP} {\bfseries 05}
  (2017) 093} [\href{https://arxiv.org/abs/1702.06539}{{\ttfamily
  1702.06539}}].

\bibitem{Alkofer:2018fxj}
N.~Alkofer and F.~Saueressig, \emph{{Asymptotically safe $f(R)$-gravity coupled
  to matter I: the polynomial case}},
  \href{https://doi.org/10.1016/j.aop.2018.07.017}{\emph{Annals Phys.}
  {\bfseries 396} (2018) 173}
  [\href{https://arxiv.org/abs/1802.00498}{{\ttfamily 1802.00498}}].

\bibitem{Wetterich:2019zdo}
C.~Wetterich and M.~Yamada, \emph{{Variable Planck mass from the gauge
  invariant flow equation}},
  \href{https://doi.org/10.1103/PhysRevD.100.066017}{\emph{Phys. Rev. D}
  {\bfseries 100} (2019) 066017}
  [\href{https://arxiv.org/abs/1906.01721}{{\ttfamily 1906.01721}}].

\bibitem{Burger:2019upn}
B.~B\"urger, J.M.~Pawlowski, M.~Reichert and B.-J.~Schaefer, \emph{{Curvature
  dependence of quantum gravity with scalars}},
  \href{https://arxiv.org/abs/1912.01624}{{\ttfamily 1912.01624}}.

\bibitem{Daas:2020dyo}
J.~Daas, W.~Oosters, F.~Saueressig and J.~Wang, \emph{{Asymptotically safe
  gravity with fermions}},
  \href{https://doi.org/10.1016/j.physletb.2020.135775}{\emph{Phys. Lett. B}
  {\bfseries 809} (2020) 135775}
  [\href{https://arxiv.org/abs/2005.12356}{{\ttfamily 2005.12356}}].

\bibitem{Daas:2021abx}
J.~Daas, W.~Oosters, F.~Saueressig and J.~Wang, \emph{{Asymptotically Safe
  Gravity-Fermion Systems on Curved Backgrounds}},
  \href{https://doi.org/10.3390/universe7080306}{\emph{Universe} {\bfseries 7}
  (2021) 306} [\href{https://arxiv.org/abs/2107.01071}{{\ttfamily
  2107.01071}}].

\bibitem{Eichhorn:2016vvy}
A.~Eichhorn and S.~Lippoldt, \emph{{Quantum gravity and Standard-Model-like
  fermions}}, \href{https://doi.org/10.1016/j.physletb.2017.01.064}{\emph{Phys.
  Lett. B} {\bfseries 767} (2017) 142}
  [\href{https://arxiv.org/abs/1611.05878}{{\ttfamily 1611.05878}}].

\bibitem{Platania:2020knd}
A.~Platania and C.~Wetterich, \emph{{Non-perturbative unitarity and fictitious
  ghosts in quantum gravity}},
  \href{https://doi.org/10.1016/j.physletb.2020.135911}{\emph{Phys. Lett. B}
  {\bfseries 811} (2020) 135911}
  [\href{https://arxiv.org/abs/2009.06637}{{\ttfamily 2009.06637}}].

\end{thebibliography}\endgroup
\bibliographystyle{JHEP.bst}
\end{document}